# Reinforcement Learning for an Efficient and Effective Malware Investigation during Cyber Incident Response

Dipo Dunsin *, Mohamed Chahine Ghanem, Karim Ouazzane, Vassil Vassilev,

***Abstract*—The ever-escalating prevalence of malware is a serious cybersecurity threat, often requiring advanced post-incident forensic investigation techniques. This paper proposes a framework to enhance malware forensics by leveraging reinforcement learning (RL). The approach combines heuristic and signature-based methods, supported by RL through a unified MDP model, which breaks down malware analysis into distinct states and actions. This optimisation enhances the identification and classification of malware variants. The framework employs Q-learning and other techniques to boost the speed and accuracy of detecting new and unknown malware, outperforming traditional methods. We tested the experimental framework across multiple virtual environments infected with various malware types. The RL agent collected forensic evidence and improved its performance through Q-tables and temporal difference learning. The epsilon-greedy exploration strategy, in conjunction with Q-learning updates, effectively facilitated transitions. The learning rate depended on the complexity of the MDP environment: higher in simpler ones for quicker convergence and lower in more complex ones for stability. This RL-enhanced model significantly reduced the time required for post-incident malware investigations, achieving a high accuracy rate of 94% in identifying malware. These results indicate RL's potential to revolutionise post-incident forensics investigations in cybersecurity. Future work will incorporate more advanced RL algorithms and large language models (LLMs) to further enhance the effectiveness of malware forensic analysis.**

## I. Introduction

Malware, also known as malicious software, is a type of software that infiltrates and compromises data and information, performing harmful and unauthorised functions. Its presence can lead to severe consequences, such as data theft, information destruction, extortion, and the crippling of organisational systems. In today's digital landscape, investigating malware has become an urgent and paramount concern due to its potential for significant damage and loss. Recent studies reveal a startling reality: malicious software is proliferating at an alarming rate, with some strains employing deceptive tactics to evade cyber forensics investigations. According to Quertier et al., [1], AV-TEST estimates the daily discovery of approximately 450,000 new malware instances, 93 percent of which are Windows-based malicious files, primarily in the form of portable executable (PE) files. This underscores the critical need for swift malware investigations when an attack occurs to prevent widespread damage and mitigate the risk of malware evolving into more sophisticated and destructive forms. Quertier et al., [1] also describe various malware investigation approaches, demonstrate the use of machine learning in malware analysis, and highlight how reinforcement learning improves malware models' performance and accuracy.

The heuristic-based malware technique is a widely employed approach that analyses various system files, categorising them as normal, unusual, or potentially harmful. Aslan and Samet [2], emphasise that while signature-based methods struggle with new malware, a combination of heuristic and signature-based approaches offers a reliable and expeditious means of identifying malicious software. In contrast, deep learning-based approaches exhibit remarkable capabilities in identifying both known and previously unseen malicious software, surpassing the performance of behaviour-based and cloud-based techniques. Malware analysis and classification heavily rely on machine learning algorithms trained to distinguish between malware and benign files. Machine learning-based approaches, according to Akhtar [3], face several challenges, including the frequent return of false positives and the ability of new malware with polymorphic traits to alter their file signatures and evade identification. To overcome these challenges, machine learning techniques train algorithms to identify and classify various forms of malware based on patterns and features found in extensive databases. These machine learning approaches improve computer systems, post-incident forensics, and network security by equipping algorithms to effectively identify and classify different types of malware [4].

While many malware instances exhibit distinct features and static structures that differentiate them from benign software, some possess characteristics that make them challenging to identify accurately [5]. Even with advances in machine learning, complex, evolving malware can evade these models and remain hidden, particularly when it is novel or highly adaptable. Reinforcement learning becomes a valuable tool in these scenarios, as it enables the creation of new malware samples capable of evading machine learning identification.

* Dipo Dunsin is the corresponding author.
Mr. D. Dunsin is with the Cyber Security Research Centre, London Metropolitan University, London, UK. email: d.dunsin@londonmey.ac.uk
Dr. M.C. Ghanem is with the Cyber Security Research Centre, London Metropolitan University, London, & Department of Computer Science, University of Liverpool, Liverpool, UK. email: m.ghanem@londonmet.ac.uk
Prof. K. Ouazzane is with the Cyber Security Research Centre, London Metropolitan University, London, UK. email:k.ouazzane@londonmet.ac.uk
Prof. V. Vassilev is with the Cyber Security Research Centre, London Metropolitan University, London, UK. email:v.vassilev@londonmet.ac.uk

These new samples retrain the malware models to identify more unknown threats [6]. Reinforcement learning distinguishes itself from conventional machine learning models by embracing uncertainty and extensive trial and error as opposed to predefined mappings [7]. This quality makes reinforcement learning particularly effective in situations where specific answers are elusive, such as in the analysis of new and unknown malware threats [8]. During the implementation of reinforcement learning algorithms, exploration plays a pivotal role. Through exploration, the model actively explores various features, expanding the breadth of knowledge about different malware types. Subsequently, exploitation comes into play, enhancing the model's performance by selecting the most beneficial attributes [9].

In reinforcement learning, the reward techniques set it apart from other machine learning approaches. Reinforcement learning provides the agent with either negative or positive evaluation feedback, which may not necessarily indicate the correct actions in the environment. Generally, we depict the agent as capable of choosing a specific set of features that, when applied, enhance the model's accuracy. The ever-changing environment, shaped by the agent's actions, facilitates the collection of relevant features for classification. According to Fang et al., [10], 'the accuracy of the classifier serves as a reward,' with the DQFSA architecture being a noteworthy example that employs reinforcement learning for feature selection. Reinforcement learning typically involves an agent that interacts with the malware analysis environment, introducing modifications to files to relate them to expected performance outcomes. According to Quertier et al., [1] recent studies such as REINFORCE and Deep Q-Network (DQN) have used reinforcement learning to improve malware investigations by leveraging past knowledge. This is particularly advantageous, as traditional machine learning models often lack the ability to incorporate background knowledge into their malware analysis. Reinforcement algorithms can reduce trial-and-error efforts and rely on past experiences to analyse and classify malware more efficiently using verified knowledge [23] & [7].

### A. Research Aim

This research seeks to enhance the accuracy and efficiency of malware forensics investigations by leveraging reinforcement learning (RL) techniques. Specifically, it aims to develop and refine models that can improve the analysis and identification of malware during post-incident investigations. The goal is to reduce investigation times and contribute to more reliable cybersecurity measures by addressing gaps in current forensics processes, particularly the limitations of heuristic and signature-based methods. Ultimately, this work strives to mitigate instances of forensic errors, such as the *'miscarriage of justice'*, and help maintain integrity in the UK's justice system.

### B. Research Objectives

The objectives of this research are fourfold. First, it aims to explore Reinforcement Learning (RL) methodologies by examining RL-based approaches for automating key tasks typically performed by forensic experts, particularly in identifying malware artefacts following a security breach. Next, the research seeks to develop a sophisticated RL model that not only classifies malware accurately but also adapts to emerging malware variants, thereby improving the efficiency of post-incident forensic analysis. Additionally, we intend to create a multi-approach malware forensics framework by integrating RL with heuristic and signature-based methods, creating a comprehensive tool for enhancing detection and analysis during post-incident investigations. Finally, we will validate the framework using real-world empirical data to evaluate its effectiveness against traditional malware forensics techniques.

### C. Research Contributions

This study makes a significant contribution to the fields of cybersecurity and digital forensics through several key advancements. First, we developed a comprehensive malware forensics approach that combines static analysis, behavioural analysis, and machine learning techniques to significantly enhance the detection and investigation of malware in memory dumps. In addition, we created a unified Markov Decision Process (MDP) model, which combines multiple MDP environments to facilitate a more structured examination of malware artefacts. Furthermore, we advanced the state of reinforcement learning (RL) in malware forensics by implementing a framework that surpasses both human and automated methods, resulting in faster and more accurate post-incident investigations while consuming fewer resources. Finally, we developed a novel method that leverages the AWK module and Volatility 3 to retrieve and identify crucial information from memory dumps, enabling a deeper understanding of malicious activities. These contributions collectively push the boundaries of malware investigation techniques.

### D. Comparison with Existing Literature Reviews

In our research, we implemented Q-learning within a defined set of action and state spaces to train reinforcement learning agents for investigating malware in post-incident malware forensics. We improve the agent's performance by continuously providing it with feedback on its performance in the MDP environment. Similarly, Binxiang, Gang, and Ruoying [11] leveraged deep reinforcement learning to overcome the limitations of traditional signature-based methods, using Q-learning to adapt to evolving malware threats, demonstrating the superiority of RL over static approaches. In a similar approach, Fang et al., [10] extended this by proposing the DQEAF model, which uses deep Q-networks to evade anti-malware engines, emphasising the creation of evasive malware that bypasses traditional malware analysis methods. The Markov Decision Process (MDP) model helped us organise our research even more. It helped us compile a list of the states and actions that make up the proposed reinforcement learning post-incident malware investigation framework. For instance, it assisted us in acquiring live memory images and identifying the operating systems in use. In a related study, Quertier et al., [1], used the DQN and REINFORCE algorithms in an MDP framework to test machine learning-based malware analysis engines and find actions that could turn malware into undetected files. In defining our action and state space, we identified ***67*** unique states and up to ***10*** actions within our RL model, facilitating a thorough malware forensics investigation. This detailed mapping is similar to the work of Ebrahimi et al., [12], who used action and state spaces in their AMG-VAC model to improve static malware analysis in black-box attack scenarios. Their study showed the pros and cons of using separate action spaces for various malware identification needs.

Additionally, we integrated various machine learning techniques, including static and behavioural analysis, to enhance our proposed framework's robustness. This integration was adapted from Wu et al. works [6] and [41], which emphasised the enhancement of malware analysis models using reinforcement learning by incorporating past knowledge into RL algorithms to improve malware identification and classification. In parallel, Piplai et al., [7] also explored the use of knowledge graphs to inform RL algorithms for malware identification, highlighting the benefits of incorporating historical data into machine learning processes. In addition, we evaluated the performance of the RL model in our research methodology based on its ability to reduce the time required for post-incident malware forensic investigations and its accuracy in identifying malware. We measure this by conducting extensive experimental testing in simulated real-world scenarios. Similarly, in the broader literature, performance metrics often include the accuracy of malware identification and the time efficiency of the forensic process. For example, a study by Raff et al., [13] evaluates their RL-based malware identification system on similar parameters, emphasising the efficiency of the RL agent in real-time scenarios. While the related work provides a solid foundation in malware analysis and MDP modelling, our research methodology builds upon this foundation by offering practical, detailed methodologies and demonstrating their application in real-world scenarios. This progression from theoretical concepts to practical implementation marks a significant contribution to the field of cybersecurity and port-incident malware forensics investigation [41] and [39].

### *E. Paper Outline*

The remainder of this paper is organised as follows: The abstract summarises the study's primary focus on leveraging RL to expedite and improve post-incident forensic processes. The introduction sets the stage by emphasizing the critical need for rapid and efficient malware investigations in light of the increasing prevalence of cyber threats, as mentioned in ***Section*** I. Following this, the literature review in ***Section*** II presents a thorough examination of existing malware analysis methods, underscoring the limitations of traditional approaches and highlighting the promise of RL. Furthermore, ***Section*** III and ***Section*** IV detail the research methodology development and implementation of the RL-based model, describing the design of the Markov decision-process (MDP) environments, the integration of reinforcement learning techniques, and the testing and evaluation of the RL agent in ***Section*** V. Furthermore, the discussion and results in ***Section*** VI show and explain the experimental results, demonstrating that the RL model works faster and more accurately than traditional methods and human forensics experts.

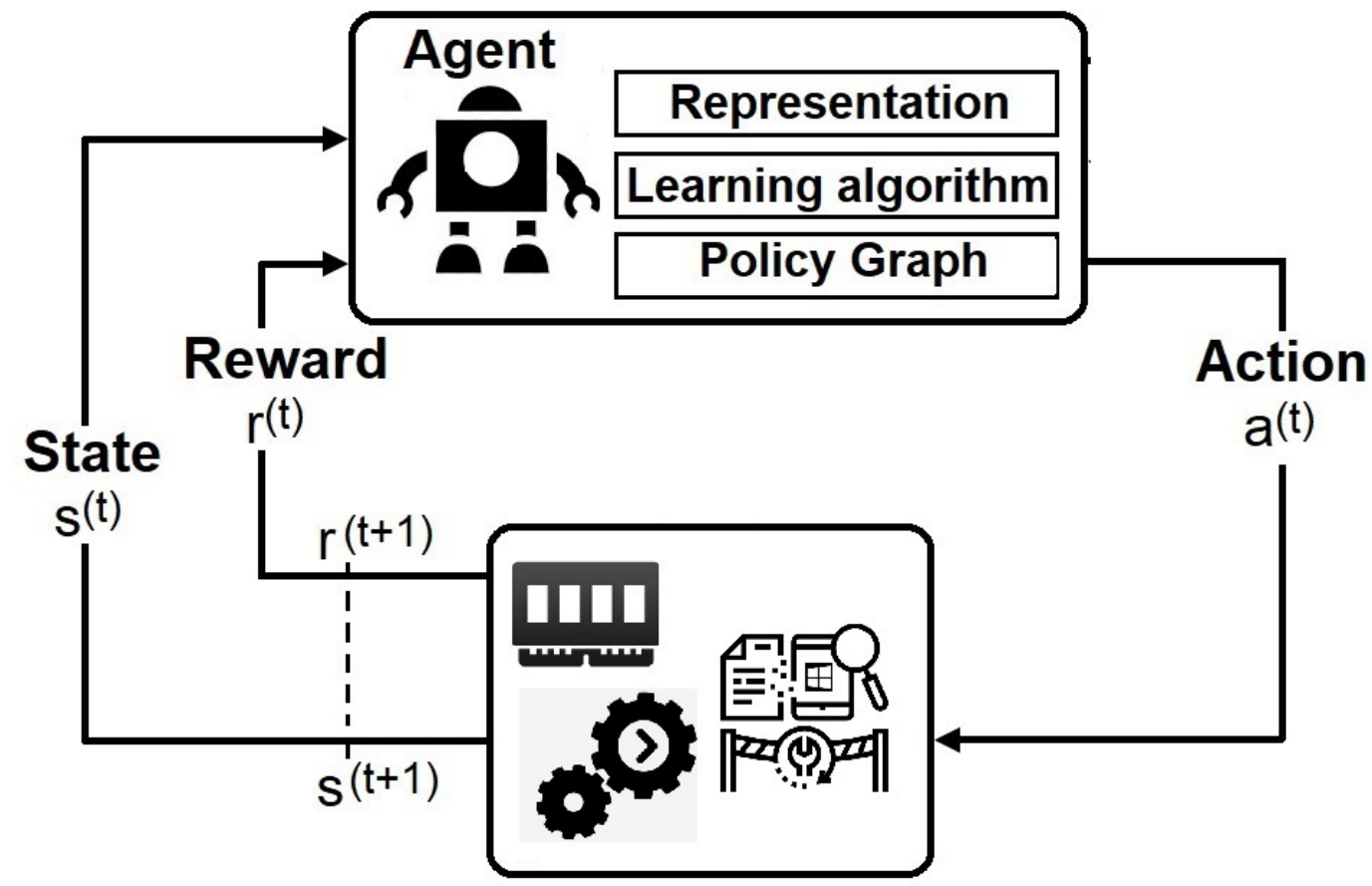

**Fig. 1:** *Malware Forensics RL Environment*

Moreover, the conclusion ***Section*** **??** encapsulates the key findings, reiterating the study's significant contributions to cybersecurity and malware forensics. It emphasises the potential of RL to revolutionise post-incident investigations, providing faster and more accurate results. Additionally, the research offers a comparative analysis, highlighting the advantages of RL over heuristic and signature-based methods. Notably, the hybrid approach integrating heuristic and RL methods shows promising results. Finally, the paper suggests some artificial intelligence techniques for future research, such as exploring advanced RL techniques and refining hybrid models, while emphasising the need for continuous learning and adaptation in RL models. The comprehensive references section supports the study, citing relevant literature on RL, malware analysis, and cybersecurity, thus providing a solid foundation for the research.

## II. Literature Review and Background

### *A. Reinforcement Learning for Malware Analysis*

Quertier et al., [1] research highlights the challenges of machine learning classifiers in identifying potential malware, especially when there is limited insight into the malware output. The study suggests that to test how well EMBER and MalConv machine learning analysis work on commercial antivirus software, one should use reinforcement learning with the REINFORCE and DQN algorithms. The study found that REINFORCE has a higher evasion rate and better performance than DQN, especially when tested against a commercial antivirus. However, a more comprehensive approach could have included training these models on a broader array of diverse models.

**TABLE I:** *Summary of Related Works.*

| Reference | Contribution | Benefits | Drawbacks |
|---|---|---|---|
| Ebrahimi et al., [12] | Proposed an AMG detector against black-box attacks | RL-based models improve malware analysis against evasion tactics. | Discrete actions may not suit all detectors, limiting universality. |
| Birman et al., [21] | Introduces real-time malware analysis using deep RL. | Enhances computer security with effective malware analysis. | Requires a large amount of training data. |
| Fang et al., [10] | Develops DQFSA for automated malware classification with deep Q-learning. | DQFSA streamlines feature selection using RL, saving time and effort. | Lacks details on restrictions imposed on the AI agent in the action space. |
| Anderson et al., [18] | Uses RL innovatively to evade PE-based malware models. | Strengthens ML-based malware analysis against adversarial threats. | Focuses on static PE models, ignoring dynamic and behavioural analysis. |
| Wu et al., [6] | Trains RL agents for optimal malware investigation models. | Improves the accuracy of ML-based malware investigations. | The need for extensive training data may hinder practical use. |
| Song et al., [17] | Proposes Mab-Malware, an RL framework for evading static classifiers. | Aids in examining evasive malware samples. | Potential misuse by malicious actors complicates analysis. |
| Rakhsha et al., [30] | Presents a practical environment poisoning algorithm for RL. | Helps develop better defenses against poisoning attacks. | Does not thoroughly address all aspects of poisoning attacks. |

### *B. Deep RL for Malware Analysis*

In 2019, Binxiang, Gang, and Ruoying [11], introduced a deep reinforcement learning-based technique for malware identification, aiming to address the vulnerabilities of traditional signature-based and machine learning-based approaches. The research demonstrated that deep reinforcement learning outperformed traditional methods based on static signatures and demonstrated the ability to quickly adapt to the ever-changing landscape of malware. However, the study had limitations, including a lack of comprehensive details about the experimental design, datasets used, and evaluation metrics. Expanding the training dataset and incorporating domain-specific knowledge could improve malware analysis. Notably, expanding the training dataset's size and diversity, as suggested by Szegedy et al., [15], could significantly enhance the effectiveness of the malware investigation. Researchers like Silver et al., [16], emphasise the value of incorporating domain-specific knowledge into deep reinforcement learning systems. While the authors assert their method's superiority, more comprehensive comparative analyses and statistical evidence could have provided more support for their claims.

### *C. RL-Based Attacks on Static Malware Detectors*

Ebrahimi et al. research, [12], aims to improve the effectiveness of static malware detectors in countering black-box cyberattacks. They propose using reinforcement learning (RL) to optimise the decision-making process of static malware detectors in the presence of black-box attacks. They create the Variational Actor-Critic for Discrete Adversarial Malware Generation (AMG-VAC) using discrete operations and an approximate sampling operator. They use RL to optimise the decision-making process, adjusting the neural network's weights based on the reward signal. In terms of accuracy, the RL-based AMG detector outperforms the original detector, particularly in the presence of black-box attacks. Furthermore, their findings show that the RL-based AMG detector is significantly more accurate than the original detector when it comes to black-box attacks. However, RL in discrete action spaces may not align with all types of malware detectors, and its effectiveness depends on factors like the training dataset quality and the neural network's architecture.

### *D. Malware Analysis Using Intelligent Feature Selection*

Fang et al., [10] developed a specialised architectural solution called DQFSA to address the shortcomings of traditional malware classification methods. The architecture uses deep Q-learning to identify crucial features, reducing human intervention and allowing data selection and analysis across various cases and data volumes. The methodology incorporates multi-view features, focusing on high classification accuracy during the validation phase. The key difference lies in exposing an AI agent to sample features with minimal human intervention. Experiments validated the DQFSA architecture by comparing its performance against various classifiers and related works. However, the DQFSA framework imposes restrictions on the AI agent within the action space that remains unexplored.

### *E. Modern Incident Response Enhanced by AI*

Dunsin et al., [19] present a study on the application of artificial intelligence (AI) and machine learning (ML) in digital forensics, focusing on enhancing malware investigation through innovative methodologies. The paper highlights the integration of AI and ML techniques to improve investigative precision and efficacy in digital forensics, leveraging advanced computational models to automate the investigation and analysis of cyber threats. Another focus is memory forensics, which focuses on machine learning algorithms to analyse memory dumps and malware, enhancing the reliability of forensic investigations by extracting and analysing multiple artefacts.

The study highlights the advantages of AI and ML in digital forensics, such as data mining techniques, reinforcement learning, and Markov decision process (MDP) for automated malware analysis. However, the study acknowledges challenges such as data validity, appropriate tools for memory dump retrieval, and adhering to ethical and legal standards. The study also proposes reinforcement learning, modelled as a Markov decision process (MDP), as a method for investigating malware in digital forensics. The MDP framework allows for systematic evaluation of different states and actions, facilitating the development of effective RL models for malware investigation, as illustrated in ***Figure 1***.

### *F. ML and Knowledge Based System for Malware Analysis*

Piplai et al., [7] propose a framework that uses reinforcement learning and open-source knowledge to enhance malware analysis. The framework consists of two components: reinforcement learning for malware analysis and knowledge from open sources detailing past cyberattacks. The research experiments create 99 distinct processes during data collection, enabling the model to identify new malware. In similar research, Gallant [20] conducted a malware investigation experiment in which the researchers trained and employed multiple machine learning algorithms, including Perceptrons, and rigorously tested their performance. However, Piplai et al., [7] leave unspecified aspects, such as determining which prior knowledge is relevant for new malware analysis and whether prior knowledge might introduce biases from previous cases. Despite these concerns, the framework's incorporation of prior knowledge remains valuable, guiding new models with increased efficiency and accuracy [38] and [40].

### *G. RL for Malware Investigations*

Reinforcement learning in malware forensics investigations involves an agent that seeks to optimise cumulative rewards by effectively managing the trade-off between exploration and exploitation. Our research's primary focus is on using reinforcement learning techniques to automate the process of conducting post-incident malware forensics investigations following a security incident. The agent performs actions within the environment, leading to changes in its state. The objective is to gradually enhance the agent's performance, enabling it to precisely identify portable executables as either malicious or benign. In the context of the proposed post-incident malware forensics investigations, the reinforcement learning agent begins at ***state zero (0)***, takes guided actions, and receives rewards or penalties. With each action and reward, it enhances its strategy through iterative learning until it attains an optimal approach.

**TABLE II:** *Malware Names Against the Operating Systems*

| Malware Variant | Target Windows OS |
|---|---|
| 1 - WannaCry | Windows 7 Professional SP2. |
| 2 - Cerber | Windows 7 All in One AIO (32/64-bit). |
| 3 - DeriaLock | Windows 11 ISO 22H2 - 64bit. |
| 4 - LuckyLcoker | Windows 8.1 Pro 6.3.9600 - 64bit |
| 5 - Dharma | Windows 8.1 Pro 6.3.9600 - 32bit. |
| 6 - SporaRansomware | Windows 10 2022 - 32bit. |
| 7 - GandCrab | Windows 10 2022 - 64bit. |
| 8 - GoldenEye | Windows 10-64bit. |
| 9 - Locky.AZ | WinDev2303Eval. |
| 10 - InfinityCrypt | MSEdge – Windows 10. |
| 11 - Win32.BlackWorm | Windows 7 Professional SP1 v6.1.7601. |
| 12 - PowerLoader | Windows 7 Starter SP1 v6.1.7601. |
| 13 - W32.MyDoom.A | Windows 7 Ultimate SP1 v6.1.7601. |

## III. Research Methodology

### *A. Experimental Setup and Dataset Generation*

To implement and validate our proposed reinforcement learning malware investigation framework, we took a systematic approach to creating a comprehensive malware dataset using the ***London Metropolitan University Digital Forensics Laboratory***. First, we established thirteen virtual machines within an isolated network to ensure a secure and controlled environment for our experiments and the eduroam network. This setup was critical to preventing the spread of unintended malware and maintaining the integrity of our data collection process. Next, we uploaded *13 different ISO files*, each representing various versions of the *Windows operating system*. This diverse selection of operating systems allowed us to test our framework across a broad spectrum of environments.

Next, we introduced a variety of malware to infect each of these operating systems. We specifically chose each malware type to represent different attack vectors and behaviours, providing a robust challenge for our investigation framework. For each ***ISO*** file installed on the virtual machine, we took an initial snapshot of the environment and saved the live memory dump. Following this, we infected the virtual machine with the chosen malware and took another snapshot. This process resulted in pairs of snapshots, one uninfected and one infected, for each operating system. This methodology produced *13 RAM files* from the uninfected environments and another *13* from the infected ones. To analyse these files, we used the Volatility framework, a powerful tool for memory forensics. We manually examined both the infected and uninfected RAM files, which, as a result, enabled us to identify significant changes and behaviours indicative of malware presence. To ensure replication and verification of our procedures, we diligently documented each stage of the analysis. This documentation was critical for maintaining the integrity of our research, as well as future reference. Finally, based on our analysis of the ***26 files***, we created a detailed malware workflow diagram. This diagram mapped out the typical processes and behaviours associated with the malware samples, providing a visual and analytical aid for understanding how different malware affects system memory. This workflow diagram is a crucial component of our proposed reinforcement learning post-incident malware forensics investigation framework, serving as a foundational element for training and validating our model. The visual diagram in ***Figure 2*** illustrates the steps we took to create the dataset, outlining each component, from setting up the virtual machine to creating the malware workflow diagram.

### *B. Malware Workflow Diagram Creation*

The research methodology extends from our comprehensive experimental setup and dataset generation process to the development of a detailed malware analysis workflow diagram, as depicted in ***Figure 3***. This diagram is integral to our reinforcement learning malware investigation framework, encompassing various malware analysis techniques, including data collection, examination, and analysis. Our dataset, comprising live memory dumps from *13* different versions of Windows operating systems—both infected and uninfected—provides the foundation for this workflow. We examined these dumps to detect anomalies, indicators of compromise, and potential malware artefacts by using the Volatility framework for memory forensics. The analysis phase incorporates a diverse array of techniques such as static analysis, signature-based analysis, behavioural analysis, and machine learning algorithms. The resulting malware workflow analysis diagram not only maps out the typical processes and behaviours associated with our chosen malware samples, but it also serves as a crucial tool for improving information security and post-incident malware forensic investigations. Our structured approach rigorously trains and validates our reinforcement learning model, strengthening our malware investigation capabilities.

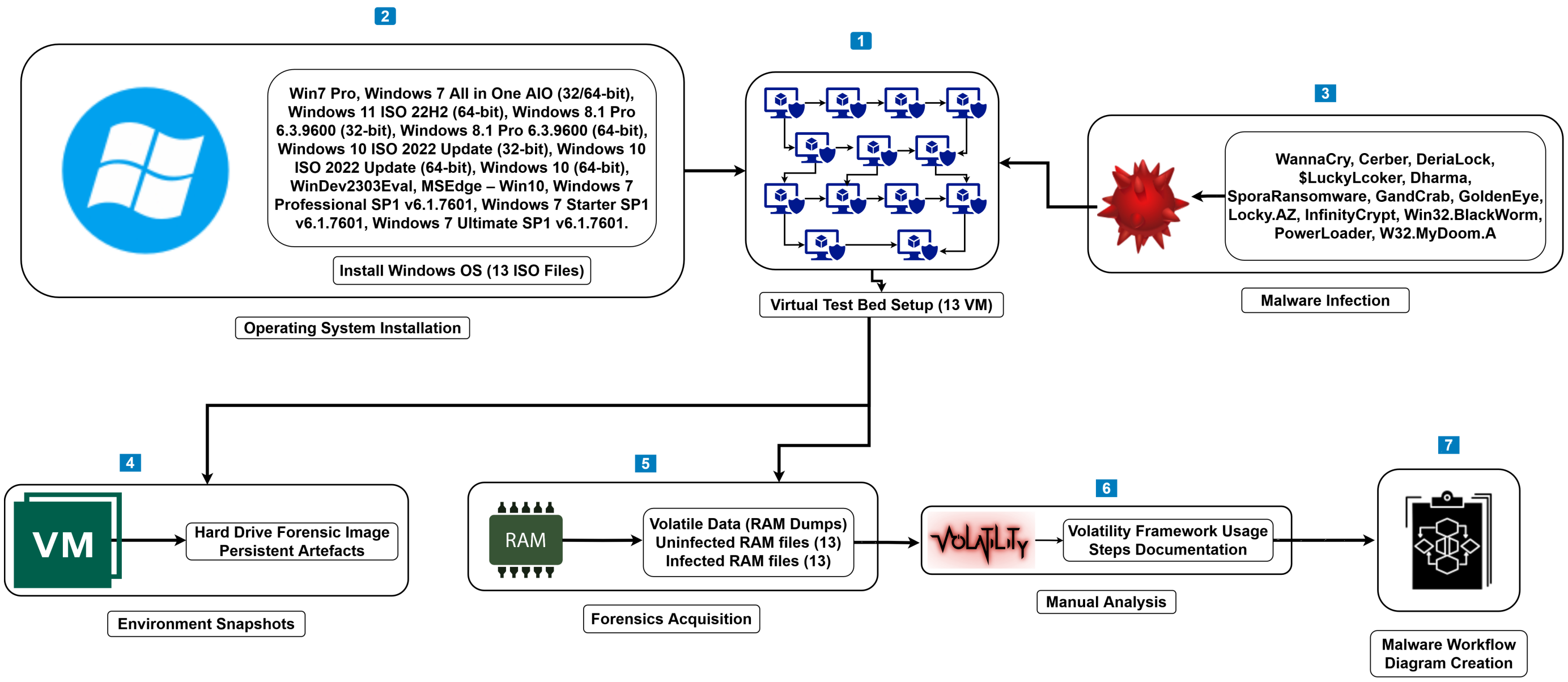

**Fig. 2:** *Experimental Setup and Dataset Generation*

### C. Markov Decision Process (MDP) Formulation

The proposed post-incident malware forensics investigation incorporates the Markov Decision Process (MDP), a mathematical framework that models decision-making when outcomes are partially random and partially under a decision-maker's control. To achieve this, our MDP consists of the following components:

- **States (S)**: In this case, $|S| = 67$ states.
- **Actions (A)**: In this case, $|A| = 10$ actions.
- **Transition Function (T)**: $T(s, a, s')$ represents the probability of transitioning from state $s$ to state $s'$ under action $a$.
- **Reward Function (R)**: $R(s, a)$ represents the immediate reward received after performing action $a$ in state $s$.
- **Discount Factor** ($\gamma$): A factor $\gamma \in [0, 1]$ that discounts future rewards.

**Step 1: Define States and Actions**

- Let $S = \{s_0, s_1, s_2, \ldots, s_{66}\}$ where each $s$ represents a unique state in the malware investigation model process.
- Let $A = \{a_0, a_1, a_2, \ldots, a_9\}$ where each $a$ represents a possible action.

**Step 2: Define Transition Function** $T(s, a, s')$

- The transition function $T(s, a, s')$ gives the probability of moving from state $s$ to state $s'$ when action $a$ is taken.
- Example: If taking action $a_2$ in state $s_5$ has a 0.8 probability of transitioning to state $s_{10}$, then $T(s_5, a_2, s_{10}) = 0.8$.

**Step 3: Define Reward Function** $R(s, a)$

- The reward function $R(s, a)$ provides the immediate reward received after taking action $a$ in state $s$.
- Example: If taking action $a_3$ in state $s_8$ gives a reward of 10, then $R(s_8, a_3) = 10$.

**Step 4: Define Discount Factor** $\gamma$

- Choose a discount factor $\gamma$ (typically between 0.9 and 1) to weigh future rewards.

### D. Leveraging Reinforcement Learning

In the context of our proposed Reinforcement Learning (RL), the agent learns the optimal policy $\pi^*$ by interacting with the three proposed MDP environments. Specifically, Q-learning, our chosen algorithm, updates the Q-values based on the Bellman equation.

*1) Value Function and Policy:* The value function for policy $\pi$ is given by:

$$V^\pi(s) = \sum_a \pi(a \mid s) \sum_{s'} T(s, a, s') \left[R(s, a) + \gamma V^\pi(s')\right]$$

- $V^{\pi}(s)$: Expected cumulative reward starting from state $s$ and following policy $\pi$.
- $\pi(a \mid s)$: Probability of taking action $a$ given state $s$ under policy $\pi$.

*2) Reinforcement Learning (RL) with Q-Learning:* subsectionLeveraging Reinforcement Learning In the context of our proposed Reinforcement Learning (RL), the agent learns the optimal policy $\pi^*$ by interacting with the three proposed MDP environments. A common algorithm used is Q-learning, which updates the Q-values based on the Bellman equation.

**The Q-learning update rule is given by**:

$$Q(s,a) \leftarrow Q(s,a) + \alpha \left[ r + \gamma \max_{a'} Q(s',a') - Q(s,a) \right]$$

**Where**:

- $\alpha$ is the learning rate.
- $r$ is the reward received after taking action $a$ in state $s$.
- $s'$ is the next state resulting from action $a$.
- $\max_{a'} Q(s',a')$ is the maximum estimated future reward from state $s'$.

***Using the specifications as a result of the workflow diagram:***

- We have 67 states and 10 actions.
- The transition and reward functions would be defined based on the specific malware identification tasks.

*3) Q-Learning Update Rule:*

$$Q(s,a) \leftarrow Q(s,a) + \alpha \left[ r + \gamma \max_{a'} Q(s',a') - Q(s,a) \right]$$

**Step 1: Initialize Q-Table**

- Initialize $Q(s,a)$ for all $s \in S$ and $a \in A$ to some arbitrary values (e.g., 0).

**Step 2: Choose Learning Rate $\alpha$ and Discount Factor $\gamma$**

- Example: $\alpha = 0.1$, $\gamma = 0.9$.

**Step 3: Implement the Q-Learning Algorithm**

- Initialize state $s$.
- Repeat:
  - Select an action $a$ using an exploration-exploitation strategy (e.g., $\epsilon$-greedy).
  - Take action $a$, observe reward $r$ and next state $s'$.
  - Update Q-Table using the Q-learning update rule:

    $$Q(s,a) \leftarrow Q(s,a) + \alpha \left[ r + \gamma \max_{a'} Q(s',a') - Q(s,a) \right]$$

  - Update state $s \leftarrow s'$.
- ***Until convergence or a specified number of episodes***.

### *E. Setting the Parameters for MDPs*

The Reinforcement Learning Post-Incident Malware Investigative Model uses the malware workflow diagram to define parameters for action and state spaces. The agent uses live memory dumps to analyse and identify malware artefacts, with ***109*** distinct actions within a defined environment. The state array aligns with the malware workflow diagram, encompassing ***67*** unique states. To achieve this alignment, we follow steps such as installing WinPmem, obtaining live memory images, understanding the operating system, extracting process information, listing DLLs, tracking open handles, collecting network data, figuring out registry hives, listing keys, duplicating processes into executable files, and sending them to Known Files Filters Servers.

### *F. The Motivation behind Implementing Q-Learning*

The proposed Reinforcement Learning Post-Incident Malware Investigation Framework uses Q-learning, an off-policy, model-free algorithm. We use it because it employs a value-based approach to determine the optimal actions based on the current state. The algorithm learns the relative value of different states and actions through experiential knowledge without relying on explicit transition or reward functions. This approach is suited for the proposed RL model for analysing malware artefacts. In this context, ***'Q'*** signifies quality, representing the action's value in terms of optimising future rewards. On the other hand, model-based algorithms employ transition and reward functions to estimate the optimal policy and construct a model, whereas model-free algorithms acquire knowledge about action outcomes experientially, without explicit transition or reward functions. In our proposed implementation, we opt for the value-based approach, which entails training the value function for the agent to learn the relative value of different states and take actions accordingly. Conversely, policy-based methods directly train the policy to determine the appropriate action for a given state. On the other hand, in off-policy methods, the algorithm assesses and improves a policy that is different from the action execution policy. In contrast, on-policy algorithms evaluate and refine the same policy employed for action execution.

### *G. Q-Learning Terminologies*

In the following sections, we will implement the proposed Reinforcement Learning Post-Incident Malware Investigation Model. The following terminologies are defined and explained in brief. An ***Environment*** is the space or world in which the agent operates and takes actions. An ***Agent*** is the entity that learns and makes decisions by interacting with the environment. ***States*** (s) signify the agent's present location within the environment. An ***Action (a)*** is the set of all possible moves or decisions the agent can make in the environment. Every action the agent takes results in either a positive reward or a penalty. ***Episodes*** mark the end of a stage, indicating that the agent cannot perform further actions. This occurs when the agent either accomplishes its objective or faces failure.

For each state-action pair, the agent uses a ***Q-Table*** to manage or store ***Q-values***. We use ***Temporal Differences (TD)*** to estimate the expected value by comparing the current state and action with the previous state and action. The ***learning rate*** is a parameter that determines how much new information overrides old information. A ***policy*** is a strategy or mapping from states to actions that defines an agent's behaviours. The ***Discount Factor*** is a parameter that determines the importance of future rewards. The ***Bellman Equation*** is a fundamental equation in Q-Learning that expresses the relationship between the Q-value of a state-action pair and the Q-values of the subsequent state-action pairs. The ***Epsilon-Greedy*** Strategy is a method for balancing *exploration* and *exploitation*.

### *H. Q-Table and Q-Function*

As previously mentioned, the *Q-table* is one of the key components that facilitate the agent's decision-making. It guides the agent in selecting the most favourable action based on expected rewards within the provided environments. The *Q-learning* algorithm updates the values of a *Q-table*, which essentially functions as a structured repository encapsulating sets of actions and states. However, defining the state and action spaces is a crucial preliminary step in effectively setting up the *Q-table*, a task that the malware workflow diagram facilitates. Furthermore, the *Q-function* plays a central role, using the *Bellman equation* and considering the *state(s)* and *action (a)* as its input. This equation significantly streamlines the calculation of both state values and state-action values.

### *I. Subsections of the Markov Decision Process Model*

The proposed subsection of the *Markov Decision Process (MDP)* represents a segment of the comprehensive and *unified MDP model*. Each subsection of the *Markov Decision Process (MDP) model* contains states, actions, rewards, and a transition probability function. These subsections are crucial components of the unified MDP that provide an agent with the capabilities of identifying and isolating suspicious portable executable files for further investigations. This approach sets a benchmark for processes such as recognising process identities, analysing process DLLs and handles, examining network artefacts, and checking for evidence of code injection.

- *WinPmem Installation*: The WinPmem MDP subset represents the different states and actions involved in the installation process, including troubleshooting and resetting if errors or corruption occur during the installation. The ultimate goal is to reach *State 5*, indicating a successful installation of WinPmem. Each action is associated with a transition between states, either progressing through the installation, returning to a previous state for debugging, or starting over. The MDP helps model and understand the decision-making process involved in WinPmem installation.
- *Acquiring Live Memory Image*: ***Figure 4*** outlines a decision-making process that involves actions, states, and transitions to achieve the goal of acquiring a live memory image, with the possibility of encountering errors and debugging them along the way. However, we structure it as a Markov decision process, where the actions taken in each state determine the state transitions. Additionally, the objective is to attain a desired state, specifically *State 10*, where successful live memory acquisition occurs.
- *Identifying the Operating System*: In this sequence, the subset MDP illustrated in ***Figure 5*** entails the transition between states and the implementation of actions aimed at identifying the operating systems. States represent the system's status, and actions are taken to achieve the goal of identifying the operating system, including debugging steps to handle errors and crashes. In *state 15*, the agent successfully identifies the suggested profile name, date, and time information using the Volatility Image Plugin, indicating successful operating system identification.
- *Identifying Process Information*: ***Figure 6*** illustrates the various states and actions available in MDP for retrieving process information. This includes using different plugin functions, debugging to fix problems, and being able to choose from different ways to collect data about the process.
- *AWK Module Features Extractions and Print List of Loaded DLLS*: The agent starts with the identified process information, executes an AWK module to extract features related to suspicious portables, and finally reaches a state where these feature extractions are complete and stored in an output file, as shown in ***Figure 7***. The next step represents a Markov decision process (MDP) for extracting information about the loaded DLLs by a specific process. The process includes executing the DLLLIST plugin, debugging potential issues, and providing the ability to reset and repeat the analysis if necessary.

### *J. The Unified Markov Decision Process*

The Unified Markov Decision Process (MDP), as depicted in ***Figure 8***, consolidates all the subsections of MDPs into a singular process, providing a comprehensive perspective. This synthesis allows the agent to effectively navigate the environment and make informed decisions regarding malware investigation.

### *K. The Proposed RL Post-Incident Malware Investigation Framework*

The Reinforcement Learning Post-Incident Malware Investigation Framework comprises six fundamental components, as illustrated in ***Figure 9***: data collection, workflow diagram mapping, MDP model implementation, environmental dependencies, MDP solver, and continuous learning and adaptation. Data collection involves acquiring live memory dumps from Windows operating systems, whereas data examination focuses on analysing the collected data to identify anomalies, compromise indicators, and potential malware artifacts. The workflow diagram outlines a comprehensive approach to identifying malware infections using static analysis, signature-based analysis, behavioural analysis, and machine learning algorithms. However, we use the AWK module to extract features from the identified processes. On the other hand, listing DLLs is an essential aspect of this workflow because it tracks loaded DLLs for each process. Additionally, monitoring open handles is crucial for keeping track of the open handles associated with each process. Another important focus is collecting network data to ensure the acquisition of all pertinent network-related information. Registry hive analysis involves identifying the registry hives and listing their keys. We duplicate the processes into executable files and check them against known malware databases to determine whether they are malicious or benign. Additionally, we duplicate the addressable memory to conduct a grep search using specific keywords.

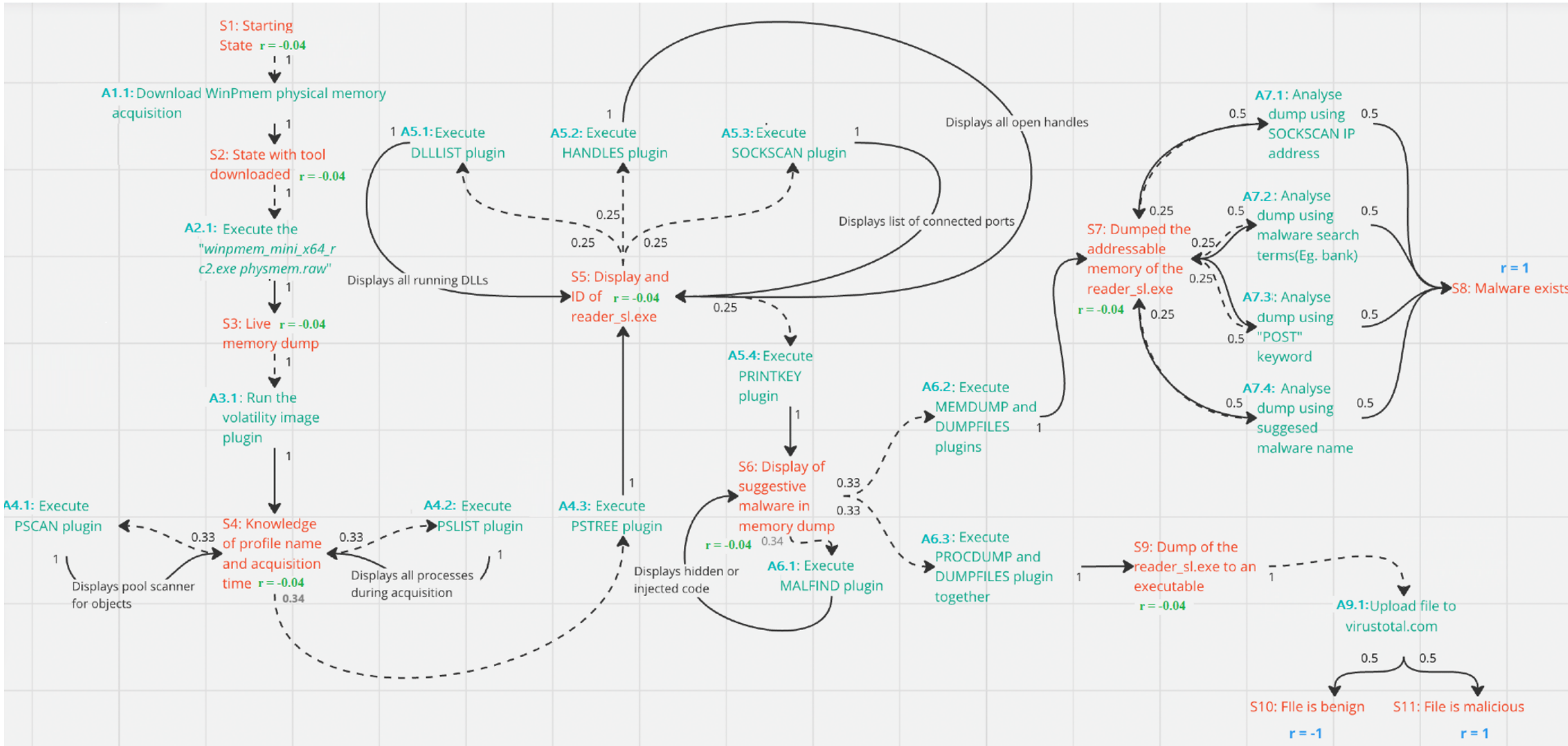

**Fig. 3:** *Malware Workflow Diagram*

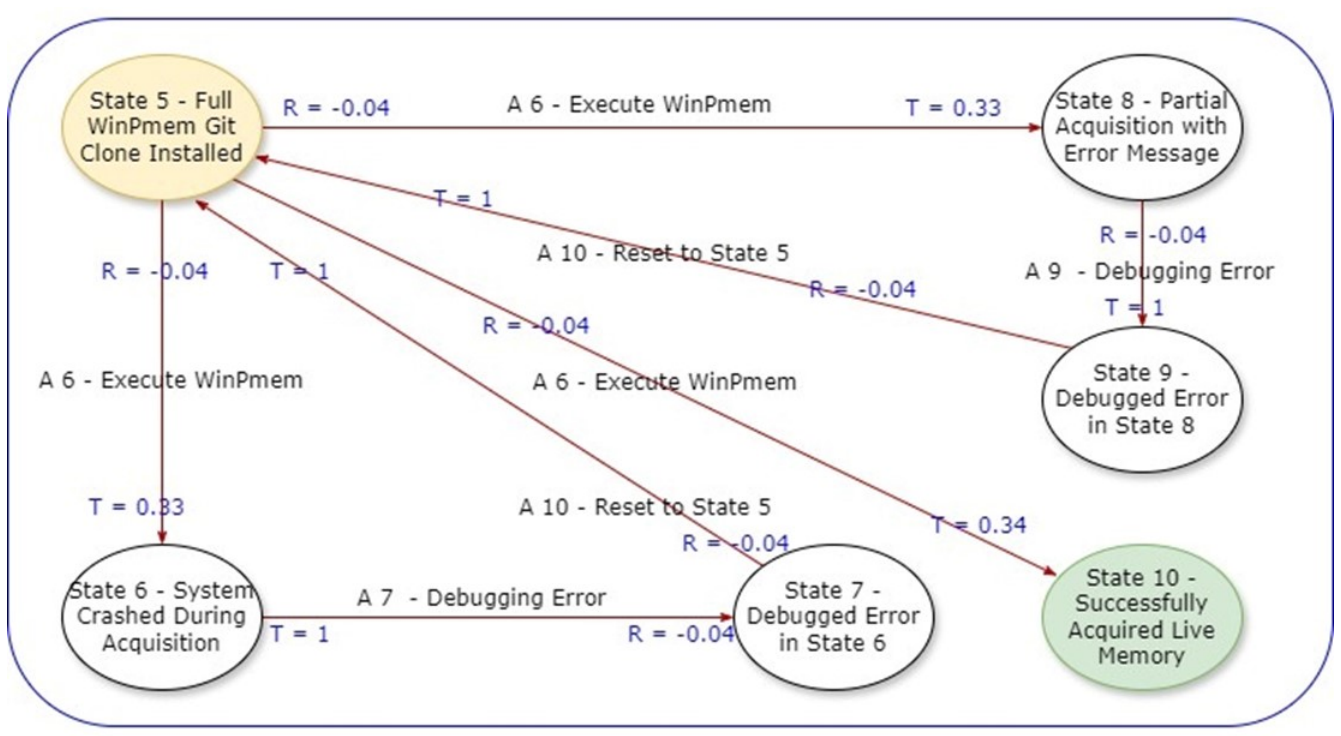

**Fig. 4:** *Live Memory Acquisition*

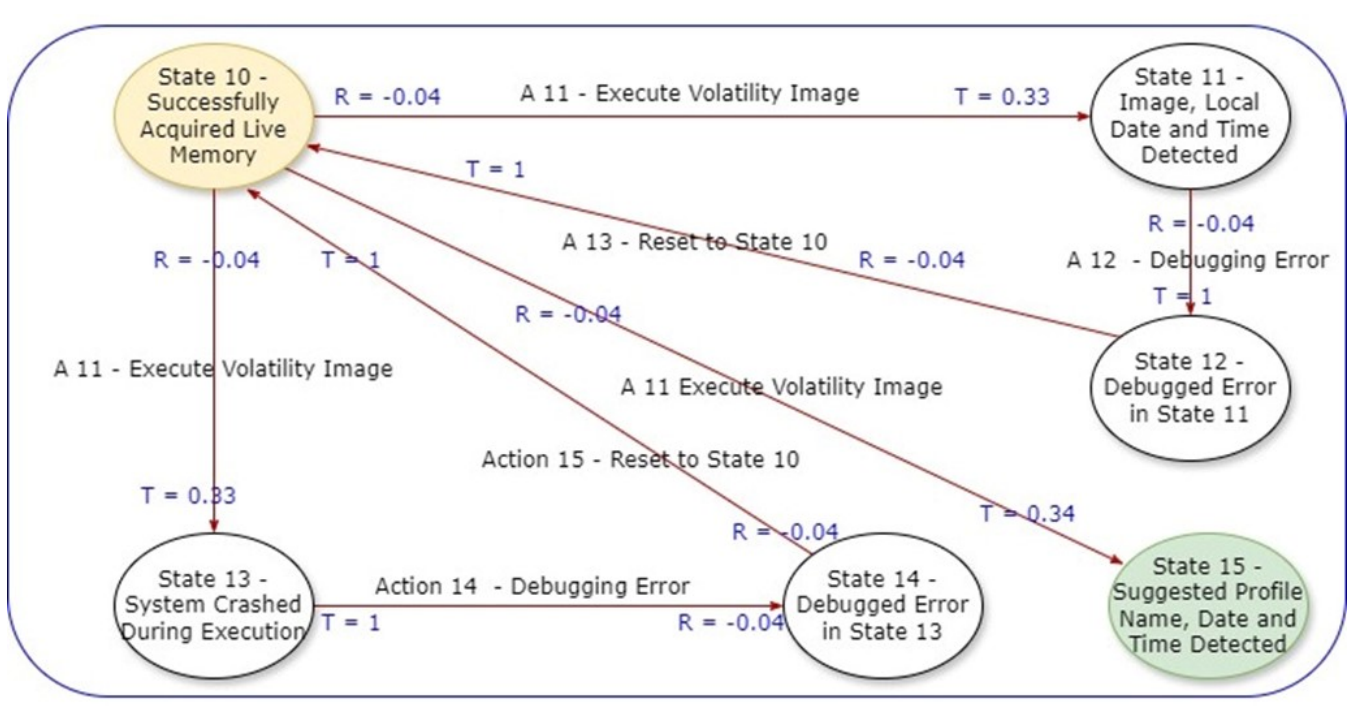

**Fig. 5:** *Identifying Operating System*

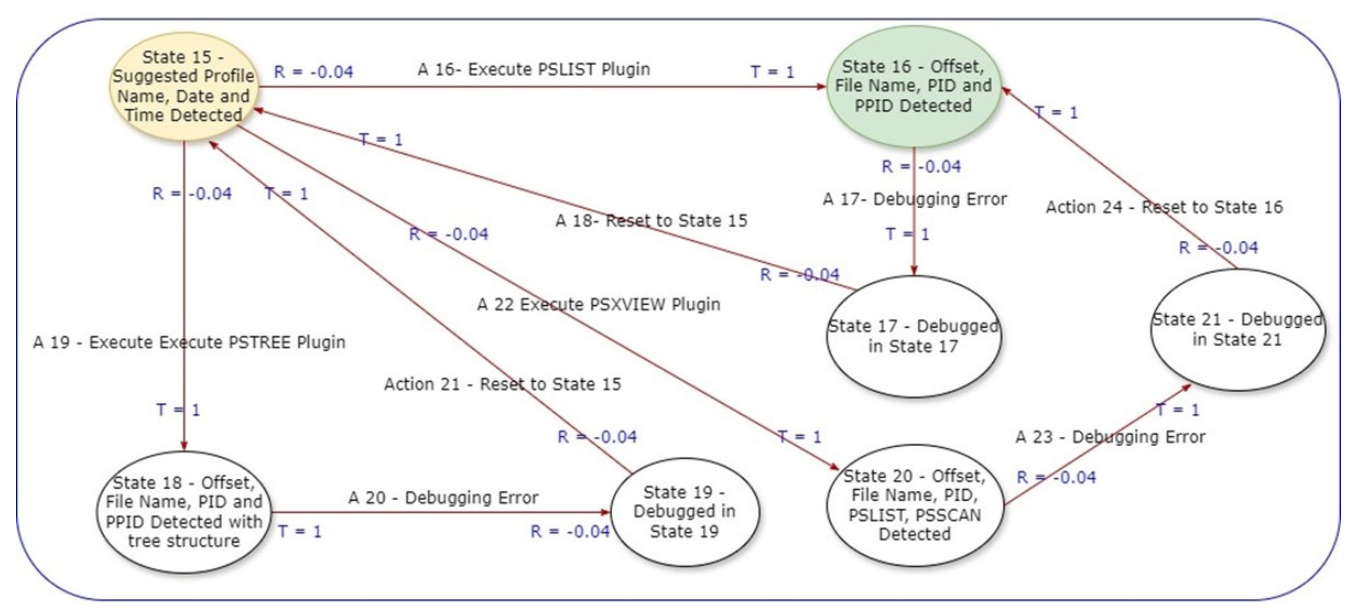

**Fig. 6:** *Windows Running Process and ID Identification*

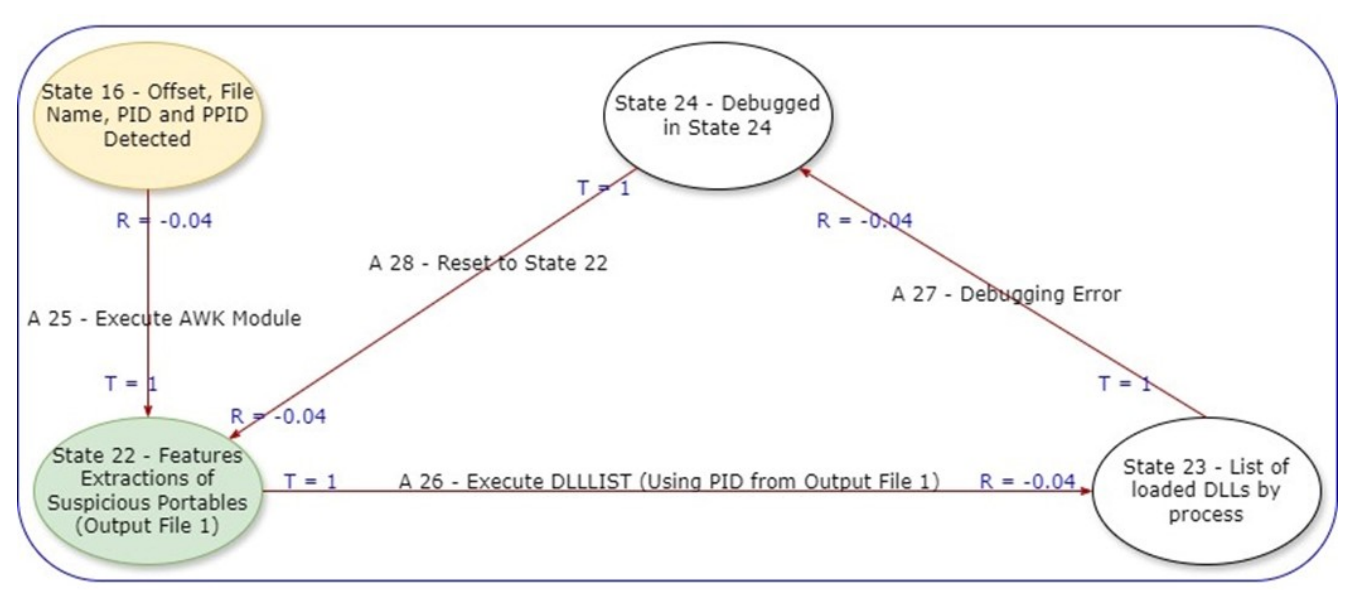

**Fig. 7:** *AWK Module Features Extractions*

The state spaces are designed to align with the malware workflow diagram, encompassing ***67*** unique states. Based on this workflow, the actions are defined, ranging from ***three*** to ***ten***, exposing the agent to ***109*** distinct actions within a defined environment. Formulating the subsections, the unified Markov Decision Process (MDP) models, and the proposed RL Post-Incident Malware Investigation Model leads to the creation of the MDP solver. We divide the setup environment dependencies into three sections: creating dependencies and gym environments, importing required libraries, and implementing the training data for continuous learning and adaptation.

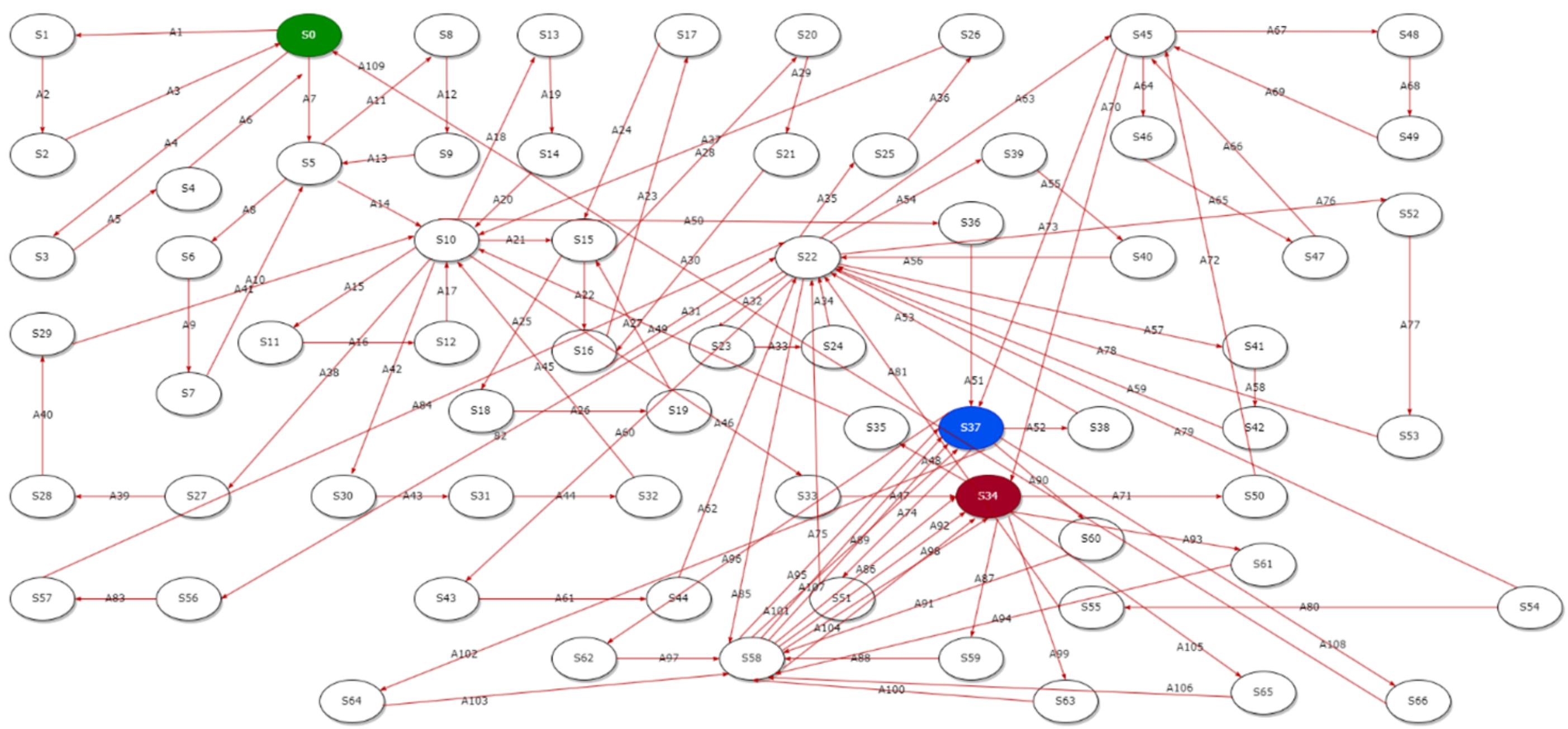

**Fig. 8:** *Overall Markov Decision Process (MDP) Model*

### *L. The Proposed RL Post-Incident Malware Investigation Model*

In the proposed Reinforcement Learning Post-Incident Malware Investigation Model, the '***Agent***' is the decision-maker which interacts with the environment. The '***Environment***' is the live memory dump in which the agent interacts. It provides the agent with state and reward data. The ***'State' (s)*** is a representation of the agent's current situation in the environment. The ***'Action' (a)*** is the set of all possible moves the agent can take. The environment provides feedback, known as the ***'Reward' (r),*** to evaluate the agent's actions. The agent uses the ***'Policy'*** as a strategy to decide the next action based on the current state. The ***'Value Function' (V(s))*** is a function that estimates the expected cumulative reward from a given state following a particular policy. The ***'Q-Function' (Q(s, a))*** is a function that estimates the expected cumulative reward from taking a particular action in a given state, following a particular policy. The agent observes the current state of the environment. The agent selects an action based on this state and its policies. The environment transitions to a new state and provides a reward to the agent. The agent updates its policy and value functions based on the reward received and the new state. This iterative process continues until the agent learns an optimal policy that maximises the cumulative reward over time.

### *M. Algorithm 1 - Implementation of the Q-Learning Algorithm*

**Algorithm 1** implements the Q-learning algorithm, a reinforcement learning technique, to train an agent to make optimal decisions in an environment. The code initially initialises a Q-table with zeros, symbolising the agent's understanding of the environment, where rows represent states and columns represent actions.

The code establishes key parameters such as the learning rate, discount factor, and exploration probability (*initially set to 0.9*), as well as decay schedules and structures for storing data (*storage, storage new, reward list*). The main loop runs for a specified number of episodes, resetting the environment and relevant variables at the start of each episode to obtain the initial state, reset the episodic reward, and initialise a step counter. A '***greedy policy***' selects actions within each episode: if the probability is high, it selects a random action (***exploration***); if not, it selects the action with the highest Q-value for the current state (***exploitation***). The environment executes the selected action, providing the next state, reward, and a done flag indicating the episode's end. The ***Bellman equation*** updates the Q-value, accounting for the immediate reward and the maximum future Q-value from the next state. The Q-table then stores the updated Q-value. Certain conditions trigger the recording of specific Q-value updates. The state is then updated to the next state, and the current and new Q-values, along with the episode number and action taken, are appended to the storage list. A predefined schedule decays the exploration probability. We optionally check convergence by comparing the absolute difference between the new and current Q-values, and if the difference falls below a threshold but the values are not equal, we can terminate the loop early. We append the episodic reward, episode number, and step count to the reward list after each episode, thereby further delaying the episode. Finally, we return the Q-table, storage, reward list, and storage new, which summaries the learned policy and the data gathered during training. The process begins with the initialisation phase, where three custom environments (*env new1, env new2,* and *env new3*) are defined using a defined *MDP function*. The algorithm then iterates over a list of names (‘*name list*’), and for each name, it assigns the appropriate environment by configuring the Markov Decision Process (MDP) with specific transition probabilities and rewards.

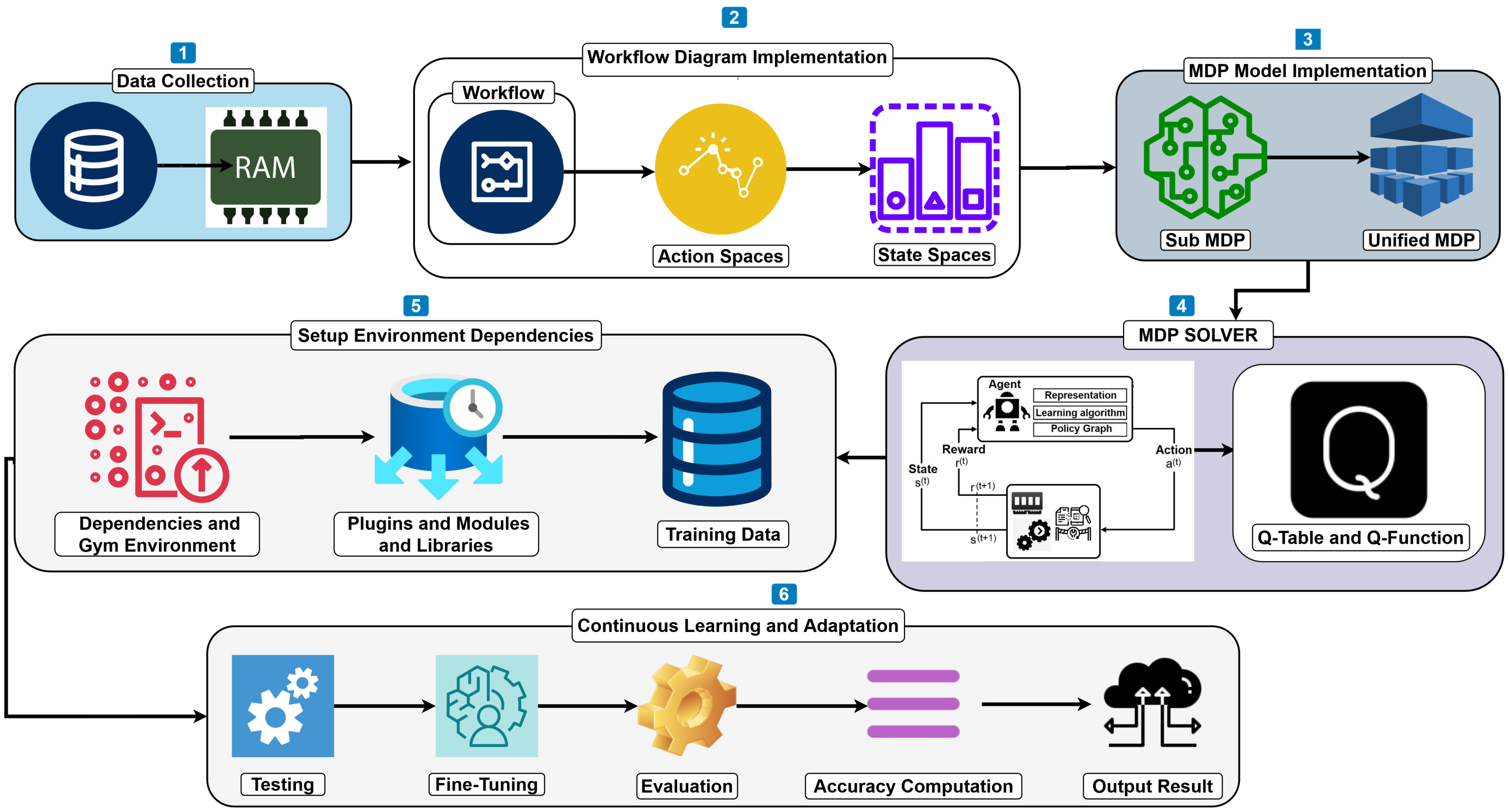

**Fig. 9:** *The Proposed RL Post-Incident Malware Investigation Framework*

### N. Algorithm -2 Iterating Learning Rates Variation over MDP Environments

**Algorithm 2** is an algorithm that *trains* and *stores* models using different *learning rates (LRs)* across multiple environments. The initialisation phase initiates the process, defining various environments (envs) and creating an empty dictionary named ’final dict’ to store results.

Next, we set the training parameters, which include a list of learning rates ranging from ‘***0.001 to 0.9***’, and store the resulting Q-tables, intermediate storage, rewards, and additional storage collections. For each learning rate (‘***lr***’) in the list of learning rates (‘***lrs***’), the algorithm executes the ‘***new q learning***’ algorithm. Finally, the algorithm stores the results by appending them to the ’***final dict***’. This structured approach guarantees systematic model training and result storage for varying learning rates in different environments.

## IV. MDP Models Integration and Implementation

### A. An Overview of Our Three Proposed MDP Environments

The ***`BlankEnvironment`*** models the proposed Markov Decision Process using the Malware Workflow Diagram, incorporating *states, actions, rewards, transition probabilities*, and *episode* completion status. It features a discrete action space with ***10*** actions and an observation space with ***67*** observations, assigning a standard step penalty of ***-0.04*** and a reward of ***2*** for identifying malware. The ***`BlankEnvironment_with_Rewards`*** gives a reward of ***2*** for all terminal states upon accurate malware identification and ***4*** for early-stage accurate identification, encouraging correct classifications. Conversely, the ***`BlankEnvironment_with_Time`*** imposes a harsher penalty of ***-0.01*** per step to incentivise efficient malware identification by discouraging the agent from taking unnecessary actions. Rewards serve as hyperparameters in both environments, refined for optimal agent performance.

### B. Implementing MDP Environments Key Python Libraries

We begin by creating the ‘***Environment***’ dependencies, importing essential libraries like *NumPy*, *Random*, *Time*, and *Gym* modules. We use NumPy for numerical computations, Random for generating random numbers, and Time for measuring code execution time. These measurements help optimise performance and compare our RL-based post-incident malware investigation model with human experts. The Gym library, commonly used in reinforcement learning, defines environments, and agents, and evaluates their performance, with its

**Algorithm 1** *Q-Learning Implementation*

1: **Initialization**:
- Initialize Q-table with zeros. (defines the state of the agent)
- Set parameters: learning rate ($\alpha$), discount factor ($\gamma$), exploration probability ($\epsilon = 0.9$), and decay schedule.
- Initialize storage structures: storage, storage_new, reward_list.

2: **for episode = 0, 1, . . . , episodes do**

3: Reset environment and variables:
- Reset the environment to obtain the initial state.
- Initialize episodic reward and step counter.
- Store current epsilon value.

4: **while not done, do**

5: Select action using $\epsilon$-greedy policy:
- if $\epsilon <$ rand() then
- action $\sim$ Uniform(noA)
- else
- action = $\max_a Q(\text{state}, a)$

6: **Execute action:**
- Act in the environment and observe the next state, reward, and done flag.
- Update episodic reward.
- Increment steps counter.

7: Update Q-value using Bellman equation

8: Compute the maximum future Q-value for the next state.

9: **Calculate the new Q-value**

10: Update the Q-table with the new Q-value.

11: Store Q-value updates if specific conditions (e.g., state, action) are met.

12: Update state: Set the current state to the next state.

13: Append the (current_q, new_q, episode, action) to the storage list.

14: Decay $\epsilon$: Reduce epsilon based on the decay schedule.

15: Check convergence: if $|\text{new_q} - \text{current_q}| <$ threshold and new_q $\neq$ current_q then

16: Break the loop

17: Append (episodic_reward, episode, steps) to reward_list.

18: Update $\epsilon$:
- $\epsilon \leftarrow \epsilon - (\epsilon\text{_decay_value} \times 0.5)$

19: Return results:
- **Return Q-table, storage, reward_list, and storage_new.**

20: **end for**

'spaces' module representing possible observations and actions in an RL environment.

## C. Initialising the State and Action Variables

We assign values to two variables: 'noS = 67' assigns the value '***67***' to the variable 'noS', implying that 'noS' represents the 'number of states,' with a value of ***67***. The variable 'noA = 10' assigns the value '***10***' to 'noA', signifying the 'number of actions', with a value of ***10***. Q-Learning will continue to use these variables to define the dimensions of data structures.

**Algorithm 2** *Iterating Learning Rates Variation over MDP Environments*

1: **Initialization**: Defining different envs and empty final dict

2: **for** name **in** name_list **do**

1) Assigning the right env (MDP): Using diff transition probs and rewards to create the MDP
2) Defining and resetting the training params:
   - Learning rates list
   - outputs, store and rewards dictionaries
3) **for** lr **in** lrs **do**
   a) Performing the new_q_learning algorithm
   b) Storing everything by appending in the final_dict
4) **end for**

3: **end for**

```python
class BlankEnvironment(gym.Env):
    def __init__(self):
        # Define action space and observation space
        self.action_space = gym.spaces.Discrete(10)
        self.observation_space = gym.spaces.Discrete(67)

        #self.action_space = np

        self.state = 0

        self.P = dict()
```

**Fig. 10:** *Initialise and Implement the BlankEnvironment Class*

## D. Implementation of the BlankEnvironment

As shown in ***Figure 10***, we define a new class named '**`BlankEnvironment`**', which inherits from '***gym.Env***', indicating its intended use as a gym environment. For the BlankEnvironment class, the constructor method initialises the class instance. The next variable defines the environment's action space and observation space. The action space is discrete, with ***10*** possible actions, whereas the observation space is discrete, with ***67*** possible observations. The next variable, '***self.state = 0***', sets the initial state of the environment to '0'. We initialize '***self.P = dict()***' as an empty dictionary to store transition probabilities in the subsequent code block.

## E. Defining Reset and Step Function

The '***def reset(self)***' function resets the environment to an initial state, returning the initial observation with '***self.state = 0***'. The '*def step(self, action)*' method simulates a step in the **`BlankEnvironment`** based on the action, assigning a random number to '*temp*' using '*np.random.rand(1)*'. If the current state is between *0* and *66* and the action is between *0* and *9*, the tuple '*k*' is updated with the current state, *1*, a reward of *-0.04*, and *False*.

## F. Implementation of the BlankEnvironment with Rewards

The **`BlankEnvironment_with_Rewards`** is a completely different implementation compared to **`BlankEnvironment`**. As illustrated in ***Figure 11***, in the **`BlankEnvironment_with_Rewards`**, actions leading

```
class BlankEnvironment_with_rewards(gym.Env):
    def __init__(self):
        # Define action space and observation space
        self.action_space = gym.spaces.Discrete(10)
        self.observation_space = gym.spaces.Discrete(67)

        #self.action_space = np

        self.state = 0

        self.P = dict()

done = k[3]
if done == True:
  reward = +4
  k = (k[0],k[1],reward,k[3])
```

**Fig. 11:** *Initialise and Implement the BlankEnvironment with Rewards Class*

to terminal states are assigned a reward of *2*, in contrast to the *-0.04* reward assigned in the ***BlankEnvironment***. The reward function in ***BlankEnvironment_with_Rewards*** is modified when an episode ends, as indicated by the done flag. The done flag assigns the value of the fourth element to the variable done, which contains information about episode completion. The done flag checks if the done variable equals True, and the reward variable is set to a positive value of *4*. This update considers the consequences of changing the reward when the episode ends.

### G. Implementation of the BlankEnvironment with Time

In ***BlankEnvironment_with_Time***, the agent incurs a more severe negative reward of ***-0.1*** per step, compared to the standard penalty of ***-0.04*** in the other two environments. This technique aims to incentivise the agent to efficiently identify malicious files by taking the most direct path, thereby discouraging any superfluous actions. Furthermore, when the agent extends the episodes by taking additional steps, it receives significant penalties. Notably, this incentive is considered a hyperparameter, as it is subject to continuous refinement. The expression '*done = k[3]*' assigns the fourth element of the tuple '*k*' to 'done', indicating whether the episode is complete. If 'done' is 'True', a reward of ***+4*** is assigned; otherwise, a penalty of ***-0.1*** is given to the agent. The tuple 'new k' is then created, maintaining the original values of '*k*' but updating the reward value. We return this updated tuple for future interactions with the environment.

### H. Iterating MDP Environments over Learning Rates

We implemented a Python code and iterated the three MDP environments over a range of learning rates (*0.001–0.9*). The `name_list = ['env_new1', 'env_new2', 'env_new3']` defines a list containing the names of the environments. We initialise an empty dictionary to store the final results and iterate over each environment using a `for` loop. We use the *Q-learning function* to convert the current learning rate to a float and store the results in dictionaries. We convert the output into a list and save it in the output dictionary. Finally, we group the collected data into a tuple and store it in the final dictionary, consolidating all results for further insight.

## V. Testing and Evaluation

### A. Retrieving Data from final-dict

We implemented a Python code and defined several variables, including Q-tables for different learning rates (`q1`, `q2`, `q3`), changes in Q-values (`store1`, `store2`, `store3`), cumulative rewards (`reward1`, `reward2`, `reward3`), and Q-values for multiple states (`store_new1`, `store_new2`, `store_new3`) for environments `env_new1`, `env_new2`, and `env_new3`. It initializes these variables by retrieving data from `final_dict`, ensuring each set of variables corresponds to a specific environment. This consistent structure allows for efficient tracking and storage of Q-learning outcomes across multiple environments.

### B. Comparing the Speed of Convergence

We implemented a Python code to visualise the speed of convergence across the three MDP environments, (`env_new1`, `env_new2`, and `env_new3`) representing `BlankEnvironment`, `BlankEnvironment_with_Rewards()`, and `BlankEnvironment_with_Time()`, respectively. Each dictionary maps learning rates to the number of episodes required for convergence. The code line `x = [float(key) for key in env_new1.keys()]` creates a list of floating-point learning rates from `env_new1`. The command `plt.figure(figsize=(10, 6))` initialises a 10x6-inch plot, where we create scatter plots for each environment, using different colours (blue, red, and green) for distinction and adding lines to illustrate convergence trends. ***Figure 12*** shows that `BlankEnvironment_with_Rewards` (`env_new2`) has the smoothest and fastest convergence. In contrast, `BlankEnvironment_with_Time` (`env_new3`) converges slowly due to a higher negative reward function, necessitating larger learning rates and more computational time. `BlankEnvironment (env_new1)` also performs well, but it converges slower due to learning rate fluctuations. As a result, `BlankEnvironment_with_Rewards` is the best MDP environment, with a 0.4 learning rate.

### C. Using Argmax to iterate over different learning rates and mdp environments

We implemented a Python code that initialises a list `lrs` with various learning rates and creates empty dictionaries `q1_dct` and `q1_dict` to store results for three different environments, `env1, env2, and env3`. The code then outputs a message indicating the processing of `env1`, then iterates over each learning rate in `lrs`, initialising lists within the dictionaries and retrieving the corresponding *Q-values* from `q1`. Within a nested loop running *67* times for different states, it prints the state index and the action index with the highest *Q-value* using `np.argmax(q_new[i])`, appending

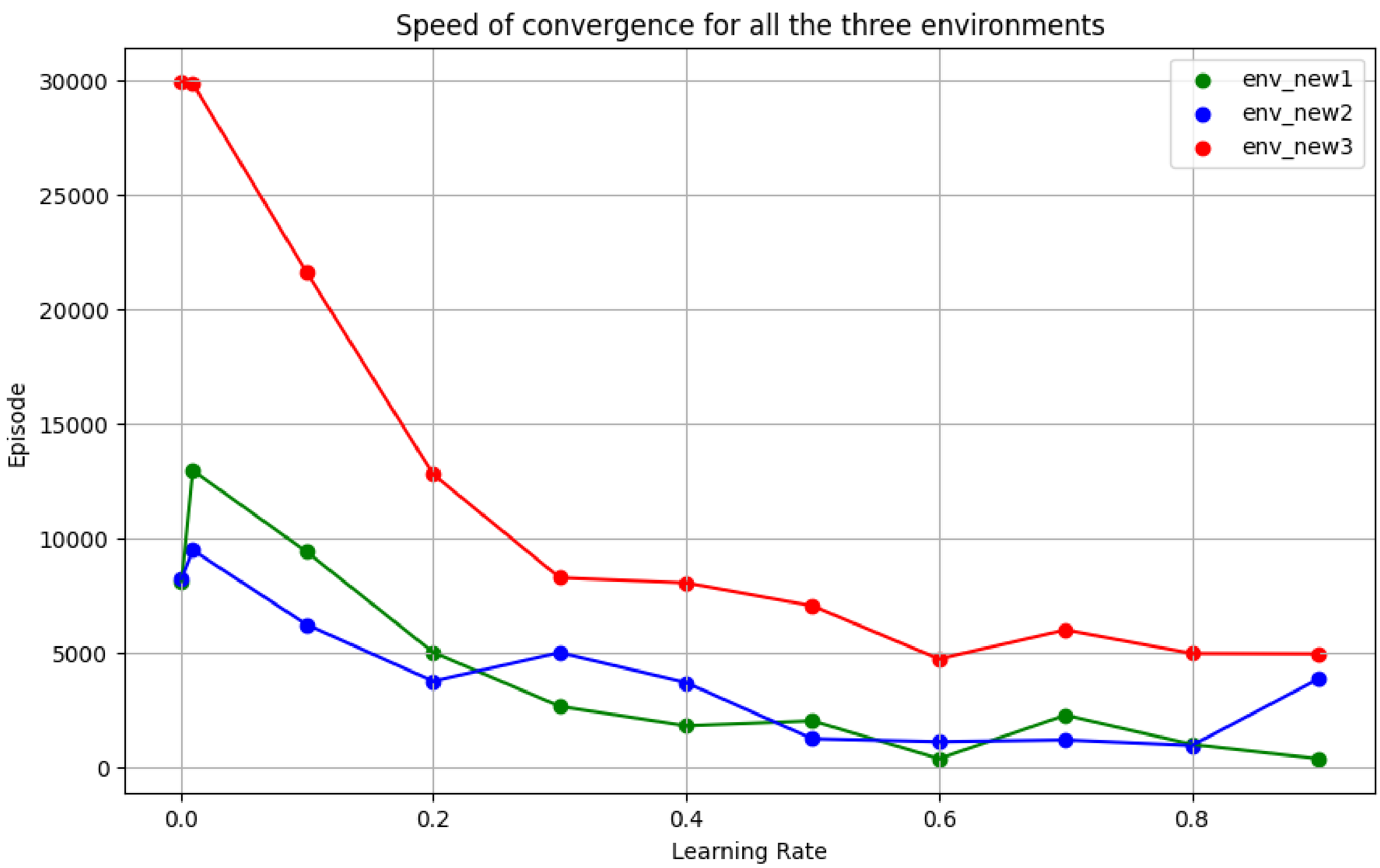

**Fig. 12:** The speed of convergence across the three MDP environments

this information as a string to `q1_dct` and as an integer to `q1_dict`.

### *D. Using Softmax to iterate over different learning rates and mdp environments*

We implemented a Python code that defines a `stable_softmax` function to calculate the softmax of an input array `x` in a numerically stable manner. It initialises a list of learning rates (*0.001–0.9*) and empty dictionaries for three environments: `env1`, `env2`, and `env3`. The code then iterates over the learning rates for each environment, processes *Q-values*, and converts them into probability distributions using the `stable_softmax` function. It then samples actions for *67* states, appending the action with the highest *Q-value* to the respective dictionary, and prints the *current environment* and *learning rate* at each step. Upon completion, it prints a `done` message, ensuring consistent performance across different environments and learning rates for further comparative analysis.

### *E. Evaluating rewards dynamics using learning rates and MDP environments*

We implemented a Python code that examines reward changes in the three *MDP environments* (`env_new1`, `env_new2`, and `env_new3`), with learning rates ranging from *0.001 to 0.9*. To create interactive plots, it imports the `plotly.graph_objects` module as `go`. The code extracts cumulative rewards, episode numbers, steps per episode, and average rewards per step from episodes *3 to 100* for each environment. We create a new figure object `fig` using Plotly and add three traces, each representing an environment with unique colours (green for `env_new1`, blue for `env_new2`, and red for `env_new3`). We update the plot layout with a centred title, axis labels, a legend title, and a hover mode that displays data for all traces at the same x-coordinate. To render and display the interactive plot, we call the `fig.show()` function, which compares the ***average rewards*** per ***episode*** for the three environments at different learning rates. However, the graph for ***learning rate 0.4*** is displayed in ***Figure 13***.

### *F. Command Definitions for State-Based Actions*

We imported the *`Subprocess`* module to allow the Python script to spawn new processes and manage their input/output/error pipes and return codes. We initialise and populate an empty dictionary, *`my_dict`*, with **key-value** pairs, each representing a ***state*** and each value a list of ***commands*** for that state. For example, *state 0* includes a command to clone a GitHub repository, while *state 10* has commands for Windows system information and the registry. *States 15 to 45* have various commands, some including special characters and options like `-pid` and `-o`.

### *G. Implementing Command Logic in Python*

We implemented a Python function *`create_command`* that takes three arguments: *`command_dict`*, *`state`*, and

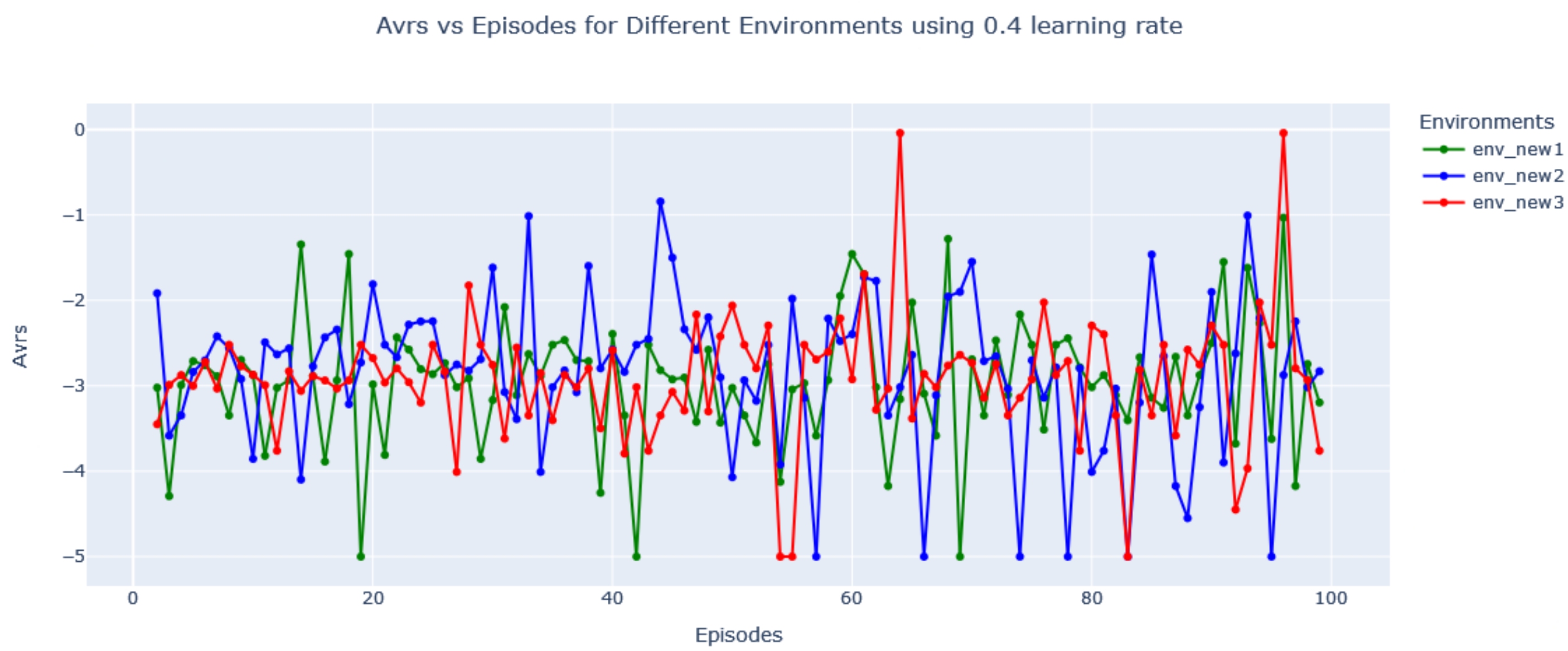

**Fig. 13:** *Average Rewards vs Episodes for Different Environment using 0.4 learning rate*

*action*. A dictionary (*command_dict*) is used to create a command based on the given *state* and *action*. The function initialises a variable *pid* with the value *340* and uses the process ID, *pid*, in the command generation logic. The *if state == 1* block checks if the state parameter is equal to *1*. If so, the code executes the next block. If the action parameter is greater than or equal to the length of the list of commands associated with the current state in *command_dict*, it is invalid. If the action is out of bounds, the code assigns the string *action out of list size* to the command variable. If the action is within bounds, the code starts an *else* block where the actual command generation takes place. The function also checks if the state variable is not in the dictionary, assigns the string *transitional state*, and checks if the action variable is greater than the list length.

### H. Defining the 'commands timings' dictionary

We set up a nested dictionary called `commands_timings` to organise the different commands that were run on the Google Collaborative Environment for WannaCry, Cerber, and Cridex malware analysis families, along with the times at which they were run (shown by variables like *ta*, *tb*, etc.). Each malware family includes specific forensic commands for analysing aspects of system memory dumps, such as process lists (*windows.pslist*), registry scans, module analysis, and network statistics. Our research relies on these commands and timings to compare the time required by a human forensics expert with the Proposed Reinforcement Learning Post-Incident Malware Investigation Framework.

## VI. Results and Discussion

### A. The Agent Decision-Making Processes

We implemented a Python script that initialises two lists, *ideal_list* and *pred_list*, containing integer values representing ***actions*** for specific ***states*** within our reinforcement learning MDP environment. The *ideal_list* assigns optimal actions for states ***0 to 66***, while the *pred_list* contains predicted actions for the same states. For example, *state 0* has an ideal *action of 0* and a predicted *action of 2*. Each index in both lists corresponds to a specific state, facilitating the comparison of predicted actions against ideal outcomes to measure model performance across the three environments using varying learning rates.

### B. Python Function to Evaluate Predictive Model Accuracy

To compare the accuracy of predicted actions against ideal actions, we implemented a Python function named *get_acc*. Initially, the function sets two variables, *true* and *false*, to zero to count correct and incorrect predictions, respectively. It iterates through the *ideal_list* and *pred_list* simultaneously using the *zip()* method, comparing each element; if they match, it increments `true`; otherwise, it increments *false*. After the iteration, the function computes the accuracy by dividing *true* by the total number of comparisons (*true + false*), formats the result to five decimal places, and prints the ***accuracy***. This function is useful for evaluating prediction accuracy in reinforcement learning settings, and upon execution, it shows an accuracy of 94%.

### C. *get_acc* function for accuracy computation

The implemented Python function named *get_acc* processes multiple environments (`env1`, `env2`, and `env3`) represented by dictionaries (`q1_dict`, `q2_dict`, and `q3_dict`). We defines *x* and *y* coordinates for three sets of data representing different environments: `env1`, `env2`, and `env3`. Each environment's data is stored in respective lists, such as `env1_x` and `env1_y`, `env2_x` and `env2_y`, and `env3_x` and

`env3_y`. The code then creates three scatter plot traces using `go.Scatter`, specifying the data points, mode (lines and markers), names, and marker colours for each environment. A layout is defined for the plot, including a title, x-axis and y-axis labels, and hover mode configuration. A figure object is created by combining the traces and the layout, and the plot is displayed using `fig.show()`. This code effectively visualises the accuracy computation for different learning rates across *three MDP environments*, as shown in ***Figure 14***. Consequently, it demonstrates that ***env2***, with a ***learning rate of 0.4***, is the best performing environment.

### D. State transitions in our RL Post-Incident Malware Investigation Model

We implemented a Python function to simulate state transitions in the ***`BlankEnvironment_with_Rewards()`*** environment, using a ***learning rate*** of ***0.4*** and driven by *actions* and *landing states*. Starting with the initial state `i` set to *0*, the loop runs *20* iterations, printing the current state (`i`) and the associated action (*`ideal_list_new[i]`*). To determine the next state, we call the function *`return_action_state`* with `i`, *`ideal_list_new[i]`*, and *`landing_list[i]`*. If `i` reaches *66*, the loop breaks. Finally, the code prints the final state (`i`) and its corresponding action (*`ideal_list_new[i]`*) after completing the loop or breaking out due to reaching state *66*. This code structure allows for simulating and tracking state transitions based on predefined actions and conditions, providing insight into how the agent navigates through our proposed RL Post-Incident Malware Investigation Model.

### E. Plotting the proposed Model command execution timings

As a result of keeping track of state changes using our proposed reinforcement learning post-incident malware investigation model, we obtain a trajectory based on actions and landing states, which control a series of state changes in the environment. We utilised the Google Collaborative Environment's execution timings to plot the proposed model's command execution timings. To store keys and values related to states and action trajectories, we created a new *`command_timings_dict`*. We then defined new Python code to create a multi-plot figure using Plotly to analyse the execution time of different malware commands (WannaCry, Cerber, and Cridex). This code initialises a figure with three vertical subplots, each with a title and increased vertical spacing. We add line plots for WannaCry, Cerber, and Cridex to the first, second, and third subplots, respectively, ensuring each has distinct colours and markers. We update the figure's layout to set its dimensions and centre the title. We customize the X-axis labels for each subplot and label the y-axes with 'Time (seconds)'. The resulting graph is displayed using `fig.show()`, as illustrated in ***Figure 15*** below.

### F. Plotting Collab and PowerShell Environment Command Execution Timings

We created a Python script to visualise the execution timings of various commands executed on Google Collaborative environments and Fast Windows Machine. The script examined WannaCry, Cerber, and Cridex to design the malware work flow diagram. The script used the Plotly library to create a multi-subplot figure, with scatter plots for each type. The graph provided a clear comparison of execution times for the three malware types. ***Figure 16–17*** displays a graph that provides a clear visual comparison of the execution times for various commands executed on the three malware types using the Google Colab environments and Fast Windows Machine in the PowerShell.

### G. Interactive Analysis of Malware Execution Times

We implemented a Python code using Plotly for interactive plotting, which allows us to visualise the execution times of various commands performed by the agent when analysing different malware, specifically WannaCry, Cerber, and Cridex. We also used the same code to sample malware analysis execution times for commands executed in both the Google Collaborative Environment and a Fast Windows Machine in the PowerShell Environment. As a result, we initialize a Plotly figure and add three separate bar plots for each malware type, with each bar representing the command's execution time. The bars for WannaCry are blue, for Cerber they are red, and for Cridex they are green. The plot's layout is customised to include a horizontally centred title, labels for the x-axis (*Commands Executed*) and y-axis (*Time (seconds)*), and grouped bars. Using `fig.show()`, we presented a visual comparison of command execution times across different malware types and environments, as illustrated in ***Figure 18–20***.

### H. Visualising execution times across multiple machines

We implemented a Python script that uses Plotly to display the total execution times for malware analysis across various machines. We iterate over the different machines and add up the execution times for each type of malware analysed (WannaCry, Cerber, and Cridex). We present the data as scatter plots with lines and markers, with each machine represented by a correspondingly designated trace. The layout contains a centred title and marked axes to help with data understanding. As a result, the chart provides a comparison view of different machines' malware analysis execution timings, as shown in ***Figure 21***. Our proposed RL Post-Incident Malware Investigation Framework demonstrates superior performance compared to the Google Collab and Windows PowerShell environments.

### I. Comparison with Traditional Malware Detection and Recent Works

The experimental results achieved in our research clearly demonstrate the advantages of implementing reinforcement learning (RL) in post-incident malware forensics. To highlight the effectiveness of our approach, it is crucial to compare these results with traditional malware detection techniques and more recent advancements in the field. Traditional forensic methods require significant human expertise and time, especially when dealing with complex or obfuscated malware. In contrast, our RL model automates much of the investigation process, demonstrating superior robustness in identifying malware artefacts from live memory images with higher precision than existing automated systems. As shown in our results, the RL agent significantly outperformed human experts in terms of analysis time while maintaining high reliability, making it a valuable tool for post-incident malware forensic investigations.

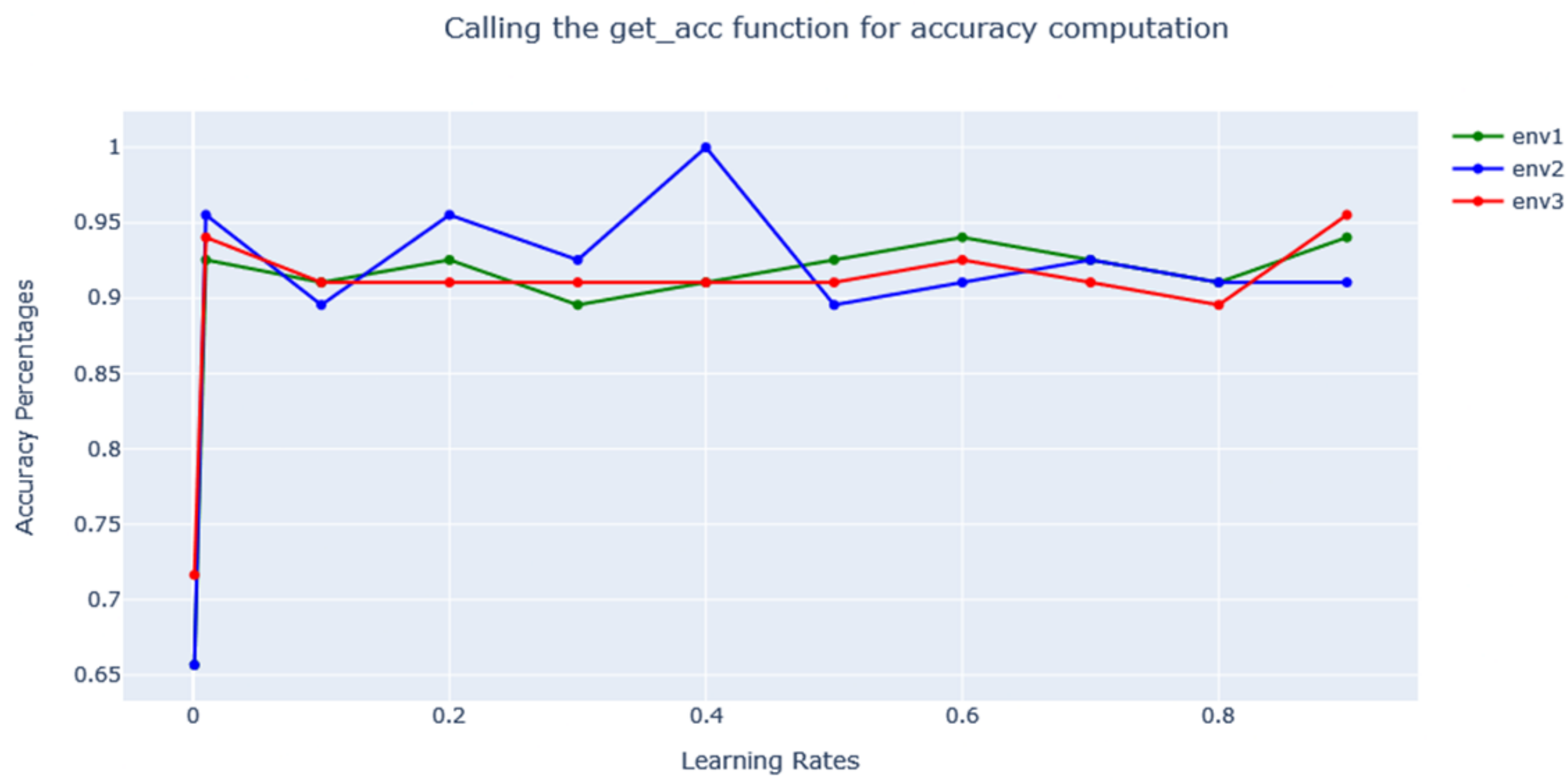

**Fig. 14:** *Calling the* `get_acc` *function for accuracy computation*

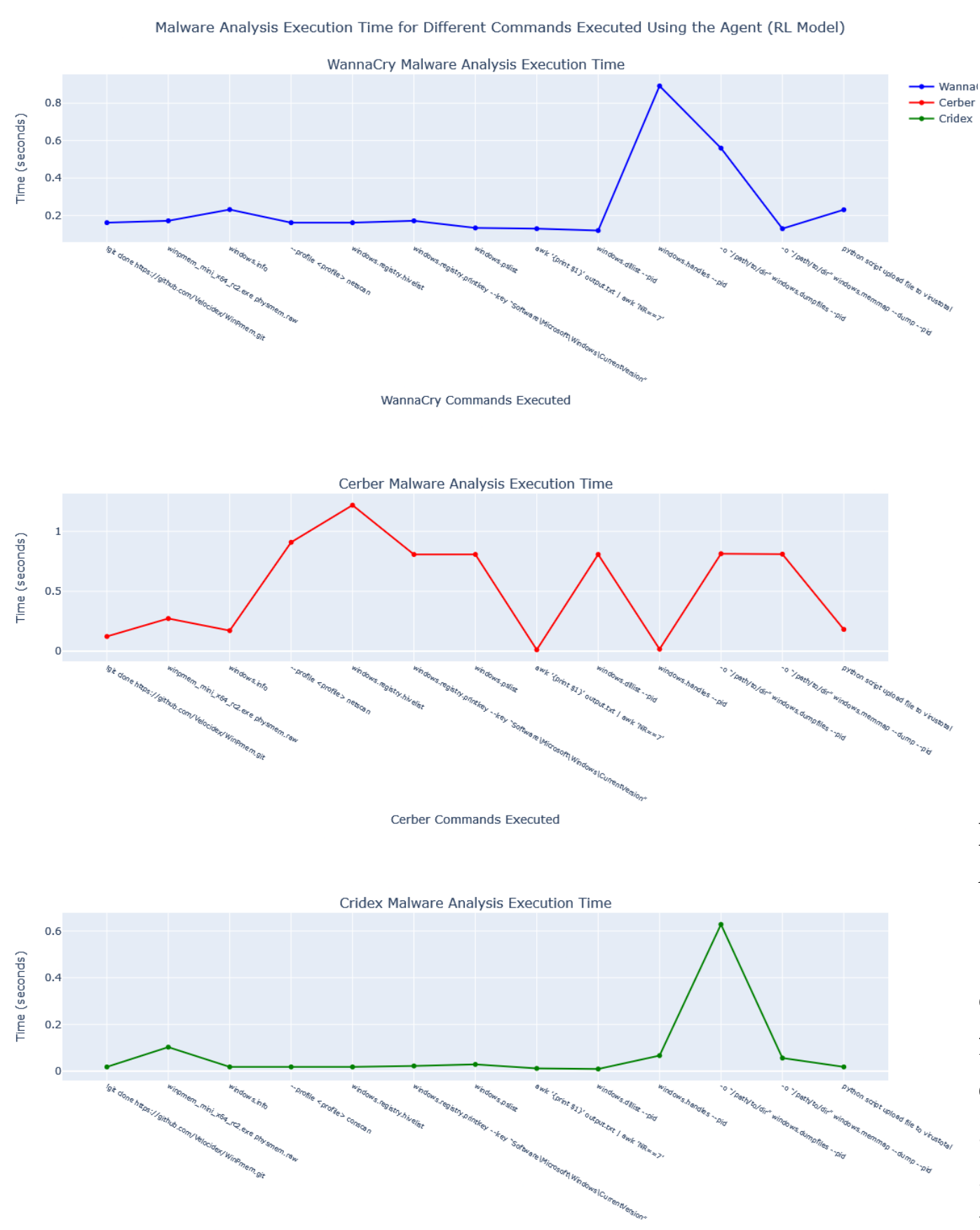

**Fig. 15:** *Malware Analysis Execution Time for Different Commands Executed Using the Agent (RL Model)*

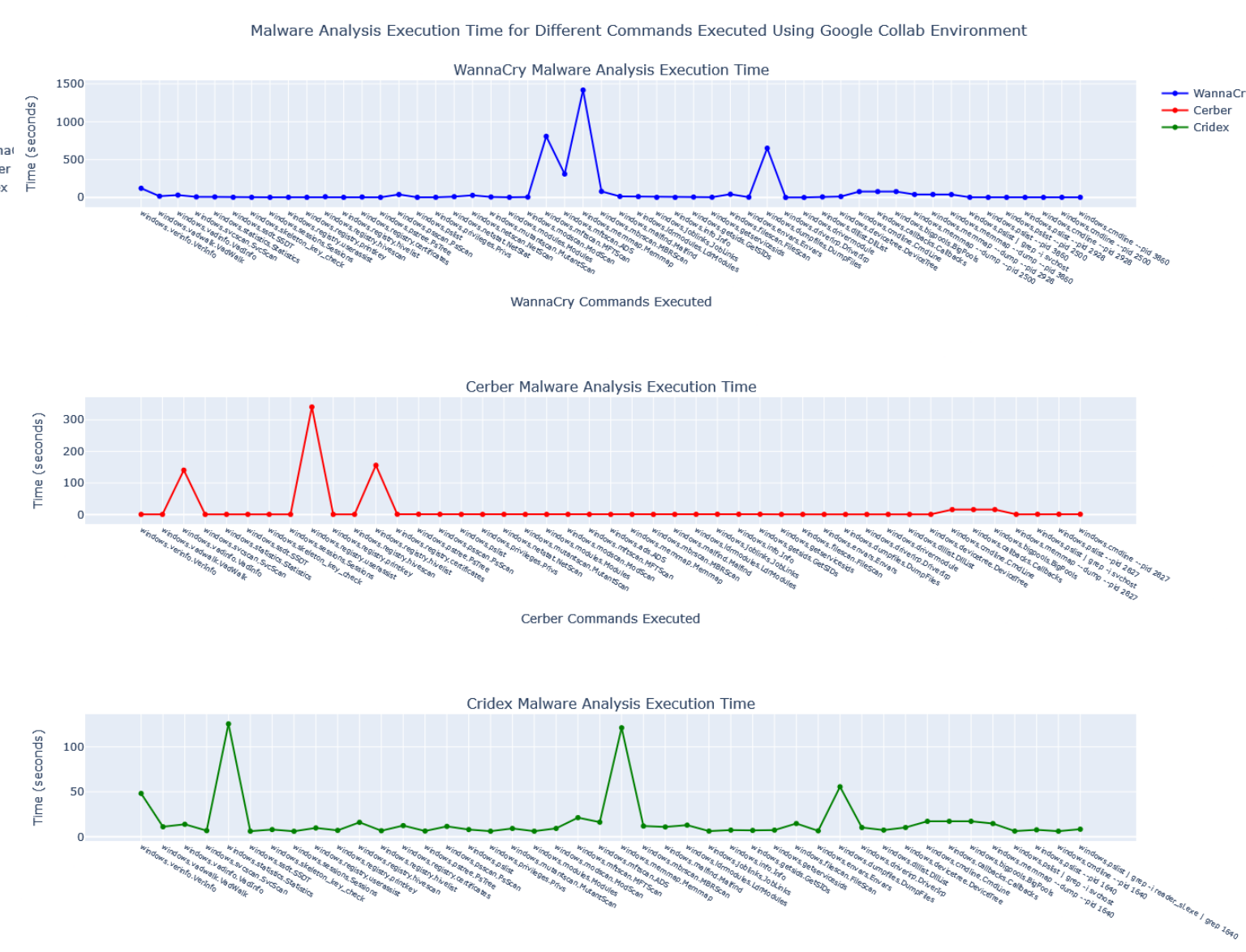

**Fig. 16:** *Malware Analysis Execution Time for Different Commands Executed Using Google Collab Environment*

According to Djenna et al., [31], traditional malware detection methods, such as signature-based and heuristic-based approaches, have been foundational in cybersecurity. Signature-based detection identifies malware by comparing files against a database of known malware signatures, which works well for detecting previously encountered threats. However, this method encounters difficulties when dealing with new or polymorphic malware, which can modify its code to evade detection. Heuristic-based methods aim to overcome this limitation by examining behaviours or patterns that indicate malware. Despite being more adaptive, heuristic methods are susceptible to high false-positive rates and still struggle with advanced, evasive malware that mimics normal system behaviour [32].

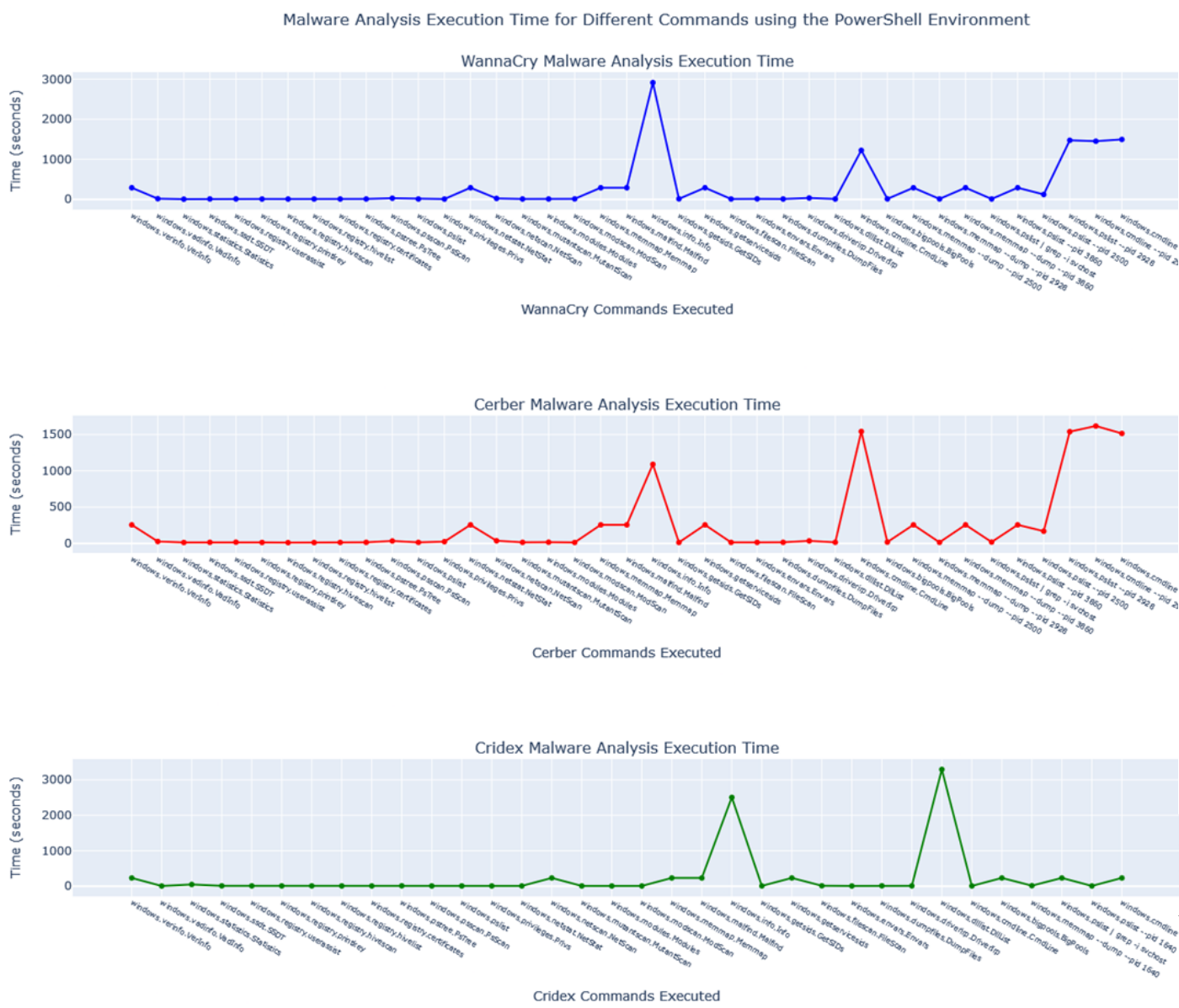

**Fig. 17:** *Malware Analysis Execution Time for Different Commands Executed using the PowerShell Environment*

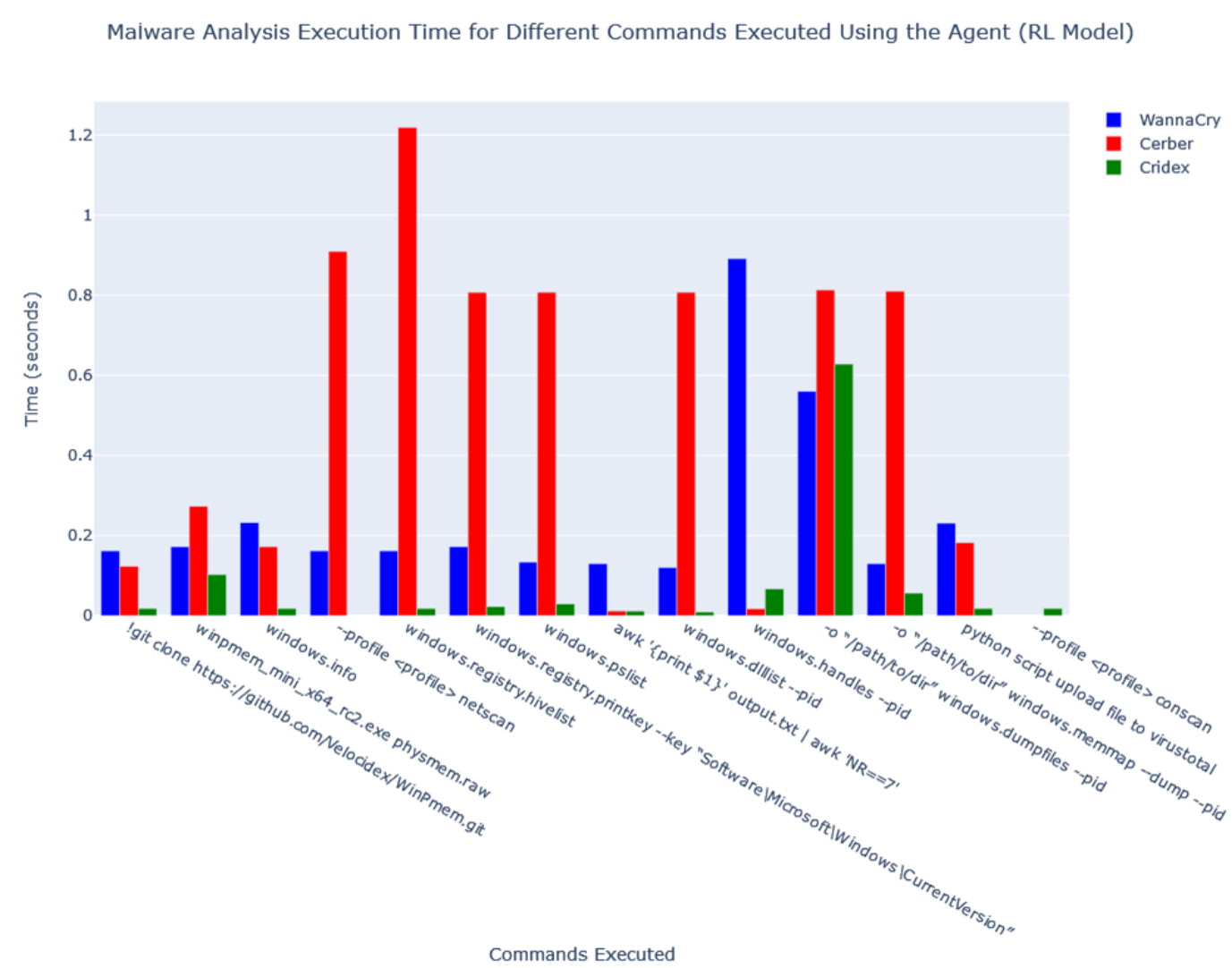

**Fig. 18:** *Malware Analysis Execution Time for Different Commands Executed Using the Agent (RL Model)*

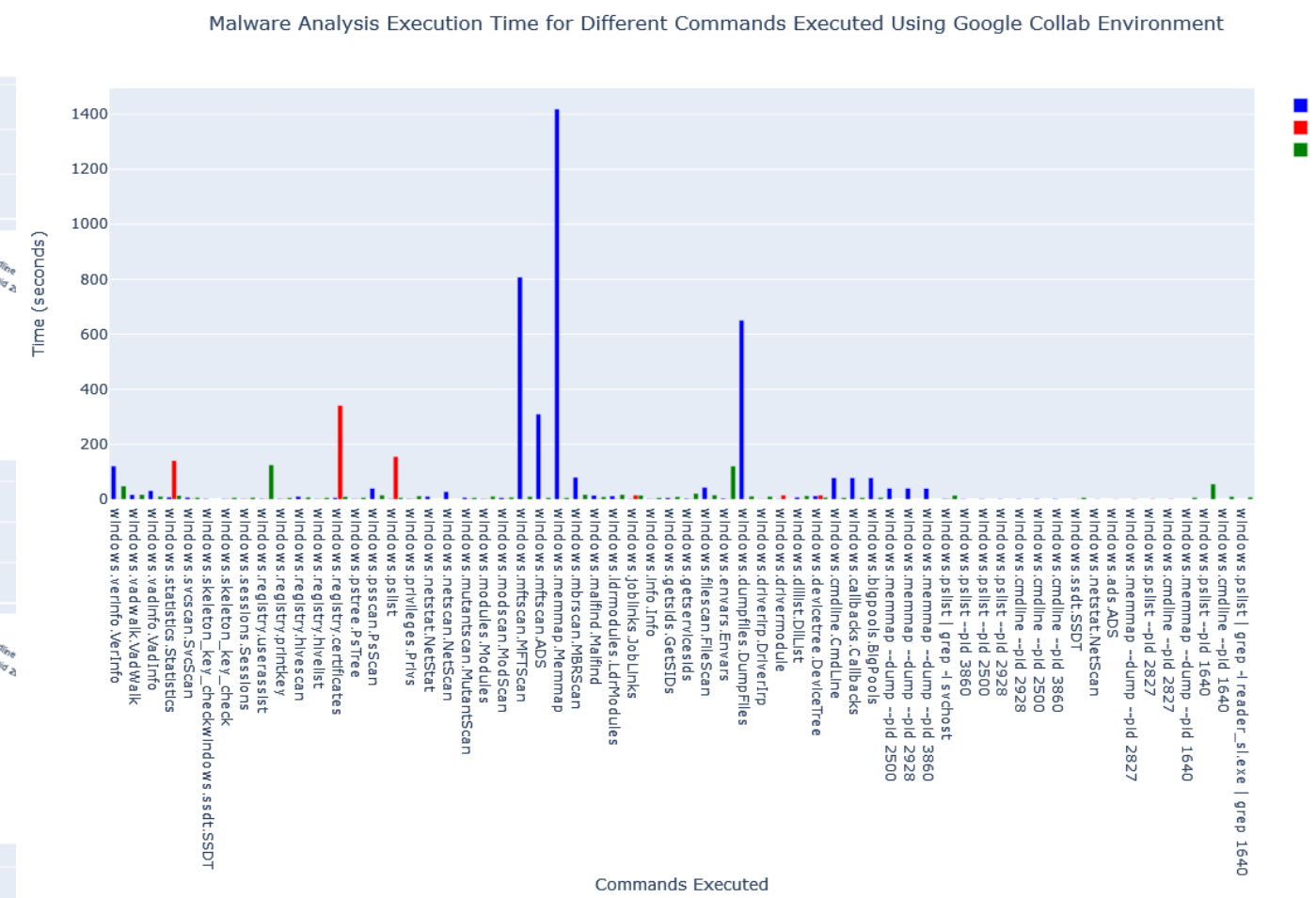

**Fig. 19:** *Malware Analysis Execution Time for Different Commands Executed Using Google Collab Environment*

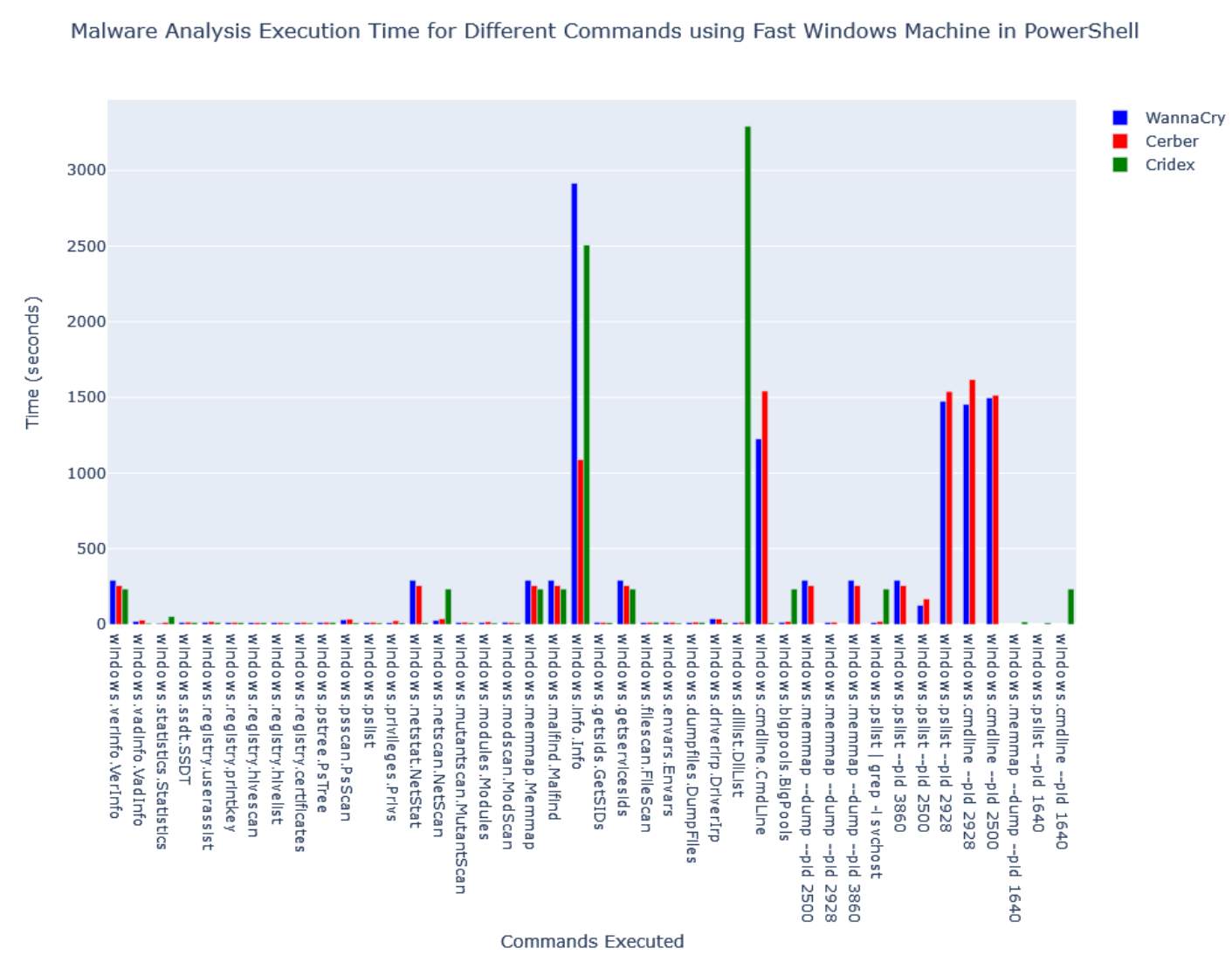

**Fig. 20:** *Malware Analysis Execution Time for Different Commands Executed Using PowerShell Environment*

Our experimental results, which used Q-learning in a reinforcement learning framework, significantly outperformed these traditional approaches. In our research, we achieved a detection accuracy of 94%, which is notably higher than what traditional methods typically reach, especially when identifying unknown malware strains. Our RL-based approach handles new and evolving malware effectively by continually learning and adapting within a Markov Decision Process (MDP) environment. This dynamic adaptability addresses the critical limitation of static signature-based and heuristic systems. The RL agent's ability to respond to a broader range of malware behaviours and features bridges gaps that traditional methods often fail to address.

Recent advancements in malware detection, particularly those involving machine learning (ML) and deep learning (DL) techniques, have sought to improve upon traditional methods [33]. Machine learning models, such as those employing support vector machines (SVMs) and neural networks, have made significant strides by enabling more flexible classification of malware based on features extracted from files or behaviours [34]. However, while machine learning approaches can better identify previously unseen malware than signature-based methods, they are often resource-intensive, requiring vast amounts of labelled data to train effectively [2]. Deep learning models, such as Convolutional Neural Networks (CNNs), have similarly demonstrated improvements in detecting malware by analysing patterns, but they tend to require large datasets and extensive computational power, making real-time detection challenging [35].

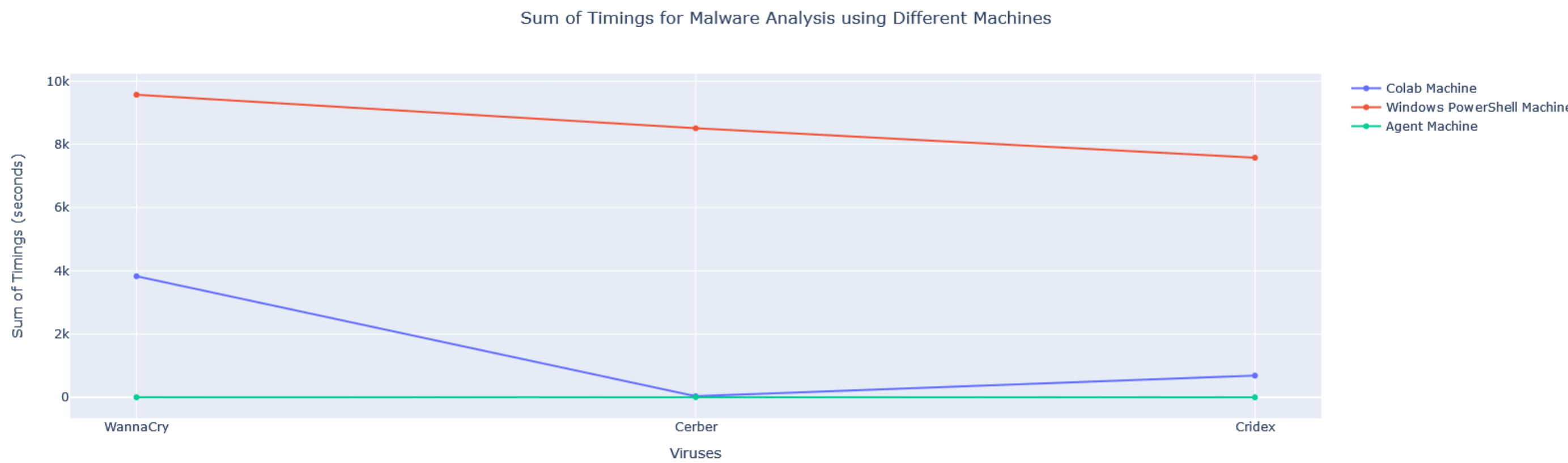

**Fig. 21:** *Total time spent on malware analysis using Collab, PowerShell, and RL Agent*

Our research builds upon these recent advancements by integrating reinforcement learning, which not only allows for learning from limited data but also enhances adaptability in real-time malware forensics. For instance, studies like Fang et al., [10] DQEAF model, which employs deep Q-networks to evade antivirus engines, emphasise the need for more dynamic approaches to deal with malware that evolves to avoid detection. In comparison, our Q-learning-based approach provides more efficient malware detection by continuously learning from its environment and refining its actions based on feedback. Our experimental results reflect this capability, as our model outperforms both traditional methods and machine learning models that rely on static data and predefined features.

Moreover, recent works, such as those by Quertier et al., [1] and Wu et al., [6] have incorporated reinforcement learning to enhance malware detection. These studies demonstrated the advantages of RL over traditional machine learning by showing how it can improve malware identification through trial-and-error learning processes. However, many of these recent works focus on specific use cases, such as static file analysis or narrowly defined malware behaviours. In contrast, our research extends the scope of RL by applying it to post-incident forensic analysis, incorporating both static and behavioural data. This broader application results in a more comprehensive and accurate detection framework, as demonstrated by our model's ability to reduce forensic investigation times while maintaining high detection accuracy.

Finally, the experimental results of our research showcase the superior effectiveness of the proposed reinforcement learning framework in comparison to both traditional malware detection methods and recent advancements in machine learning and deep learning. As a result of addressing the limitations of static, signature-based, and resource-intensive machine learning models, our RL-based framework adapts to evolving malware threats dynamically and integrates both static and behavioural analyses. This comprehensive approach makes our method a robust and efficient solution for post-incident malware forensics investigations.

### *J. Research Findings and Recommendations*

The paper proposes a post-incident malware investigation framework built upon a novel MDP model that leverages advanced reinforcement learning (RL). The model significantly speeds up the investigation process, surpassing human forensic experts in both the speed and detection of known and unknown malware threats. It integrates various malware analysis techniques and includes data collection methods like live memory dumps from Windows systems. A custom malware dataset and comprehensive malware workflow diagram were created to streamline the forensic process.

The core of the approach is a unified Markov Decision Process (MDP) model that combines multiple MDP environments into one cohesive framework. Three distinct environments were created with each employing a unique reward structure to guide the RL agent in developing optimal malware analysis strategies. The RL model operates within these structured MDP environments, allowing the agent to navigate the malware analysis workflow. Each state and action corresponds to a specific stage in the malware analysis process, enabling the agent to learn, estimate, and refine the expected value of actions. This dynamic learning is driven by the Q-learning algorithm, which balances the exploration of unknown states with the exploitation of known policies, optimizing decision-making. A Q-table manages state-action pairs, while temporal-difference learning iteratively updates the agent's knowledge base, improving malware identification accuracy over time.

Extensive experimental evaluation showed that the learning rate is key to convergence, with simpler environments benefiting from higher rates and more complex environments requiring lower rates for stability. In realistic post-incident scenarios using malware such as WannaCry, Cerber, and Cridex, the model demonstrated strong classification accuracy, adaptability to novel threats, and computational efficiency, indicating robustness and scalability. Iterative refinement of the MDP environments, guided by experimental feedback and hyperparameter tuning, was crucial to optimizing the RL agent's performance. Fine-tuning learning rates and reward mechanisms across diverse scenarios greatly enhanced the model's effectiveness. The RL-based approach for malware forensics offers a promising alternative to traditional methods, with the potential for real-time adaptability to evolving malware threats. Future research should focus on optimizing reward functions, expanding state-space designs, and integrating advanced feature extraction techniques like behavioural analysis, temporal pattern recognition, hybrid static-dynamic feature analysis, and adversarial training to further enhance the framework's applicability in dynamic forensic environments.